\documentclass[leqno,a4paper,12pt,oneside]{article}
\usepackage[latin1]{inputenc}
\usepackage[english]{babel}
\usepackage[T1]{fontenc}
\usepackage{amsfonts,amssymb,amsmath,amsthm,latexsym,amscd,mathtools,exscale}
\usepackage{makeidx}
\title{An historical prolegomenon to the occurrence of the resonance dual models: the Watson-Sommerfeld transform}
\author{\small Giuseppe Iurato\\\footnotesize\it University of Palermo, IT}
\date{}
\begin{document}\setcounter{tocdepth}{22}\maketitle\begin{abstract}We would like to discuss, within the inherent historical
context, upon one of the main works of fundamental physics which has led to the formulation of the early
resonance dual models prior to string theory. In fact, we shall focus on certain aspects of the fundamental
Tullio Regge work of 1959, to be precise, on certain mathematical methods handled by him to pursue his original
intentions mainly motivated to prove some double dispersion relations into the framework of non-relativistic
potential scattering.\end{abstract}It is well-known that the celebrated Gabriele Veneziano work of 1968 is
almost unanimously considered as the first step toward the formulation of resonance dual models of particle
physics, from which then string theory soon arose. Nevertheless that, we are of the historiographical opinion
that the early prolegomena to the first resonance dual models is the famous 1959 seminal paper of Tullio Regge,
in which, in introducing complex orbital momentua to determine further possible connections between potentials
and scattering amplitudes, a notable formal technique was introduced, namely the so-called \it Watson-Sommerfeld
transform\footnote{Often also said to be \it Watson-Sommerfeld-Regge transform.}. \rm Thenceforth, such a formal
tool even more played a fundamental and pivotal role in theoretical particle physics, so that we would like to
highlight the chief historical moments which led to such a formal technique, trying to identify the truly early
sources of it. The historical course treated by us in pursuing this end, comprises two main steps, namely a
first one having an introductory character which is devoted to a very brief outline of the main formal elements
of scattering theory, and a second one just centered on such 1959 Tullio Regge paper. In achieving this, a
special attention has been paid to those many unavoidable moments concerning history of mathematics and
mathematical-physics whose preeminent presence, along this historical route, cannot be evaded or neglected. But,
let us say it immediately: we will not deal with the wide history of scattering theory, because our main aim
just consists in directly arguing on the rising of 1959 Regge work about the introduction of complex angular
momenta and, above all, on the related mathematical methods there involved, pointing out just some historical
aspects concerning these latter but that have nevertheless been quite neglected by the usual historical
treatments of the subject.

\subsection*{1. Scattering theory: a brief general overview} Most of information concerning quantum systems comes
from collision theory. Following\footnote{This textbook is one the main references that we have taken into
account in drawing up this paper. Surely, (Rossetti, 1985) is the central reference considered in drawing up the
Section 1.} (Rossetti 1985, Capitolo I), the early crucial moments of the history of quantum theory has seen
basically involved celebrated collision experiments: to mention the main ones only, we recall the 1911 E.
Rutherford experiments on elastic collision of $\alpha$ particles on atoms, the 1913-14 J. Franck and G. Hertz
inelastic collisions of electrons on gas atoms and molecules, the photoelectric effect, the 1923 A.H. Compton
scattering of photons on atomic electrons, the 1927 C. Davisson and L. Germer diffraction experiments and the
1934 E. Fermi experiments on neutron collisions; on the other hand, the modern elementary particle physics is
also based on collision phenomena: for instance, the celebrated 1983 C. Rubbia experimental observation of
intermediate vector bosons essentially was a proton-antiproton annihilation process, which constituted one of
the first experimental confirmations of the Standard Model. In this paper, we are interested only in those
historical aspects of collision theory which have led to the dawning of resonance dual models of string theory,
with a particular attention to the many intersection points with history of mathematics.

One\marginpar{\tiny\it Scattering cross-sections} of the chief notions of collision theory for putting together
experimental data and theoretical issues, is that of \it scattering cross-section. \rm The \it differential
scattering cross-section \rm $d\sigma=\sigma(\Omega)d\Omega$ is defined to be the number of scattered particles
within the infinitesimal solid angle $d\Omega$ (with $\Omega$ computed in polar coordinates $0\leq\theta\leq\pi$
and $0\leq\varphi\leq 2\pi$) per unit of time and per unit of incident beam of target particle. The \it total
scattering cross-section \rm is defined to be
\begin{equation}\sigma_t(E)\doteq\int_{\Omega}\sigma(\Omega)d\Omega\end{equation}which depends only upon the
energy $E$ of the incident beam. Following (Landau \& Lif\v{s}its 1982, Chapter XVII) and (Rossetti 1985,
Chapter II), in the center of mass reference frame, the motion of two particles\footnote{With negligible
spin-orbit interaction.} with well-defined energy $E$ and interacting by means of a central potential field
$V(\vec{r})$, may be reduced to the motion of a single particle with reduced mass $m$ and relative energy $E$,
which is ruled by the following well-known stationary Schr\"{o}dinger equation
\begin{equation}-\frac{\hbar^2}{2m}\Delta\psi(\vec{r})+V(\vec{r})\psi(\vec{r})=E\psi(\vec{r})\end{equation}
which is the equation of motion of a particle with mass $m$ moving in a central potential field $V(\vec{r})$. If
we set $k^2=(2m/\hbar^2)E$ and $U(\vec{r})=(2m/\hbar^2)V(\vec{r})$, the equation (2) will assume the form
\begin{equation}\Delta\psi(\vec{r})+[k^2-U(\vec{r})]\psi(\vec{r})=0.\end{equation}We suppose
$U(\vec{r})\rightarrow 0$ as $\vec{r}\rightarrow\infty$ in such a rapid manner that we might speak of so afar
regions where the interaction is negligible and where the above parameter $k$ will be equal to the modulus of
the wave vector $\vec{k}$ of the relative motion.

We consider an incident beam directed along the positive $z$ axis, as polar axis, oriented according to the
related wave vector $\vec{k}$. At infinite distances, the wave function $\psi_{as}(\vec{r})$ will satisfy the
following asymptotic wave equation
\begin{equation}\Delta\psi_{as}(\vec{r})+k^2\psi_{as}(\vec{r})=0\end{equation}whose solutions are either
(progressive and regressive) plane waves of the form $\psi_{as}(\vec{r})=Ae^{\pm i\vec{k}\cdot\vec{r}}$ and
(divergent and convergent distorted) spherical waves of the form $\psi_{as}(\vec{r})=f(\theta,\varphi)e^{\pm
i\vec{k}\cdot\vec{r}}/r$, so that the wave function $\psi(\vec{r})$ describing a scattering process, at infinite
distance, is the superposition of an incident plane wave (associated to the incident beam) and of a wave
emerging from the scattering center. Therefore, we have the following asymptotic behavior
\begin{equation}\psi({\vec{r}})\underset{r\rightarrow\infty}{\sim}\psi_{as}({\vec{r}})=A\Big\{e^{i\vec{k}\cdot\vec{r}}+
f(\theta,\varphi)\frac{e^{i\vec{k}\cdot\vec{r}}}{r}\Big\}\end{equation}where $A$ is a normalizing constant,
$f(\theta,\varphi)$ is a function, called \it scattering amplitude \rm and having a length dimension, which
estimates the amplitude of the scattered distorted spherical wave with respect to the amplitude of the incident
beam, $\theta$ is the planar scattering angle of the emerging beam with respect to the $z$ axis and $k$ is the
modulus of the wave vector $\vec{k}$. The geometry of the system implies that $f(\theta,\varphi)$ depends on the
scattering angle $\theta$ (colatitude) but not on the anomaly $\varphi$. In the scattering amplitude are
included the most important information on the scattering process, whose knowledge involves exact solutions to
the original Schr\"{o}dinger equation, hence the form of $U(\vec{r})$. Furthermore, as regard the differential
scattering cross-section, it is possible to prove that
\begin{equation}\sigma(\Omega)=\sigma(\theta,\varphi)=|f(\theta,\varphi)|^2\end{equation}while, as regard the
total scattering cross-section, we have
\begin{equation}\sigma_t(k)=\int_{\Omega}|f(\theta,\varphi)|^2d\Omega,\end{equation}in the geometry of central
potentials having these formulas themselves but with $f(\theta)$ for $f(\theta,\varphi)$. Therefore, the
scattering cross-section may be estimated by means of the scattering amplitude, and vice versa.

In non-relativistic potential scattering theory, we have to consider a central short-range potential field
$U(r)$. Usually, the $z$ axis will be the axis for the scattering center, coincident with the origin of the
given polar coordinate system, parallel to the wave vector $\vec{k}=\vec{p}/\hbar$ of the incident beam.
Therefore, we have a dynamical problem having cylindric symmetry with respect to the $z$ axis, so that almost
all the most important physical quantities will be independent from the anomaly $\varphi$ (like
$f(\theta,\varphi)\rightarrow f(\theta)$). In this case, under certain amplitude normalization conditions,
equation (5) reduces to the following
\begin{equation}\psi({\vec{r}})\underset{r\rightarrow\infty}{\sim}e^{i\vec{k}\cdot\vec{r}}+f(\theta)
\frac{e^{i\vec{k}\cdot\vec{r}}}{r}\end{equation}which is a solution to the following Schr\"{o}dinger equation
\begin{equation}[\Delta+(k^2-U(r))]\psi(\vec{r})=0,\end{equation}and is the sum of two terms, an incident plane wave
and an emerging spherical wave (scattered wave) that contains the scattering amplitude from which it is possible
to deduce all the physical characteristics of the collision process. The equation (9), for central potentials,
has a set of elementary solutions given by
\begin{equation}\psi_{lm}(\vec{r})=\varphi_l(r)Y^m_l(\theta,\varphi)\end{equation}which forms a complete set of
solutions, so that every other solution to (9) will be a linear combination of them. In the geometry of the
dynamical problem that we have chosen above, only those spherical functions not depending on the anomaly
$\varphi$ may be considered, that is to say $Y^0_l(\theta,\varphi)=\sqrt{(2l+1)/4\pi}P_l(\cos\theta)$, so that
our wave function has a Legendre polynomial series expansion of the type
\begin{equation}\psi(\vec{r})=\psi(r,\theta)=\sum_{l=0}^{\infty}C_l\varphi_l(r)P_l(\cos\theta)\end{equation}with
$\varphi_l(r)$ radial wave functions which have well-defined behaviors as $r$ varies in $[0,\infty[$. In
particular, we are interested in the asymptotic behavior for sufficiently large values of $r$ in which potential
and centrifugal terms (due to finite potential barriers) are either negligible, given by
\begin{equation}\varphi_l(r)\underset{r\rightarrow\infty}{\sim}\frac{A_l}{kr}\sin\Big(kr-l\frac{\pi}{2}+
\delta_l\Big)\end{equation}where $A_l$ is a suitable coefficient which goes to 1 as $U(r)\rightarrow 0$ and
$\delta_l$ is an asymptotic phase displacement or shift of the given solution to the radial Schr\"{o}dinger
equation, related to the physical interaction and computed with respect to the free interaction solution. On the
other hand, as $U\rightarrow 0$, the wave function defined by the asymptotic formula (9) should reduce to the
free solution given by the incident plane wave, so that, taking into account (11), we will have
\begin{equation}\psi(\vec{r})\underset{U\rightarrow 0}{\longrightarrow}\psi_{free}(\vec{r})\equiv
e^{i\vec{k}\cdot\vec{r}}=e^{ikr\cos\theta}=
\sum_{l=0}^{\infty}i^l(2l+1)j_l(kr)P_l(\cos\theta)\end{equation}where $j_l(kr)$ are the Bessel spherical
functions in the argument $kr$. In turn, (11) might be written as an expansion in partial waves each having a
definite angular momentum $l$, of the type\footnote{First expansions of this type were provided by W. Gordon in
(Gordon 1928).}
\begin{equation}\psi(\vec{r})=\sum_{l=0}^{\infty}i^l(2l+1)\varphi_l(r)P_l(\cos\theta)\end{equation}where
$\varphi_l(r)$ are solutions to the radial Schr\"{o}dinger equation, regular at the origin and normalized in
order to reduce them to Bessel functions $j_l(kr)$ as $U\rightarrow 0$.

Inserting\marginpar{\tiny\it Phase shifts, partial amplitudes} (12) into (14), we have
\begin{equation}\psi(\vec{r})\underset{r\rightarrow 0}{\sim}
\sum_{l=0}^{\infty}i^l(2l+1)\frac{A_l}{kr}\sin\Big(kr-l\frac{\pi}{2}+
\delta_l\Big)P_l(\cos\theta)\end{equation}so obtaining an asymptotic partial wave development of the wave
function which basically dependent on the phase displacements, or \it scattering phases\rm, or \it phase shifts,
$\delta_l$, \rm since it is possible to prove that also the constants $A_l$ are function of the latter. In turn,
$\delta_l$ are determined by the potential $U(r)$ and vice versa: indeed, the knowledge of $\delta_l$ will allow
the reconstruction of the interaction potential (scattering phase-function method). Therefore, these latter play
a very fundamental role in potential scattering theory\footnote{Following (Landau \& Lif\v{s}its 1982, Chapter
XVII), the problem how to build up the form of scattering potential from the scattering phases supposed to be
known, has a particular relevance and has been solved by I.M. Gelfand, B.M. Levitan and V.A. Mar\v{c}enko. In
this regard, it turns out to be that, for the determination of $U(r)$, it is enough to know $\delta_0(k)$ as
function of the wave vector in the region comprised between $k=0$ and $\infty$ as well as the coefficients $a_n$
in the following asymptotic expressions$$\varphi_{l}\approx a_le^{(\sqrt{2m|E_l|}/\hbar)r}/r $$ of the radial
wave functions $\varphi_l$ (of (11)) relative to those states corresponding to negative discrete energy levels,
if these exist. The determination of $U(r)$ from these latter data involves the resolution of a certain linear
integral equation (see (De Alfaro \& Regge 1965)). Furthermore, another very interesting method to determine the
scattering phase is the so-called \it variable phase method \rm (see (Calogero 1967)).}. They have real values
for real values of the potential, whilst, for complex values of the potential, the phases have in general
complex values as well. On the other hand, as pointed out above, the scattering cross-section is closely related
with the scattering amplitude which, as now we shall see, is related with the asymptotic form of the wave
function, hence, ultimately, with the phase displacements. Thus, measuring scattering cross-sections, we shall
be able to get information about phase displacements, hence about the interaction potential. Indeed, following
(Rossetti 1985, Capitolo II), it turns out to be
\begin{equation}\begin{split}f(\theta)=&\frac{1}{2ik}\sum_{l=0}^{\infty}(2l+1)(e^{2i\delta_l}-1)P_l(\cos\theta)=\\
=&\frac{1}{k}\sum_{l=0}^{\infty}(2l+1)e^{2i\delta_l}\sin\delta_l\cdot
P_l(\cos\theta)=\\=&\sum_{l=0}^{\infty}(2l+1)a_l P_l(\cos\theta)\end{split}\end{equation}having put, as a \it
partial amplitude \rm corresponding to the partial wave of angular momentum $l$\begin{equation}a_l\equiv
a_l(k)=\frac{1}{2ik}(e^{2i\delta_l}-1)=\frac{1}{k}e^{i\delta_l}\sin\delta_l,\end{equation}so that, being
$\sigma(\theta)=|f(\theta)|^2$, we have the following expression for the differential scattering cross-section
\begin{equation}\sigma(\theta)=\frac{1}{k^2}\Big|\sum_{l=0}^{\infty}(2l+1)
e^{2i\delta_l}\sin\delta_l\cdot P_l(\cos\theta)\Big|^2\end{equation}whilst for the (elastic) total scattering
cross-section we have
\begin{equation}\sigma_{tot}(k)=2\pi\int_{-1}^1\sigma(\theta)d\cos\theta=\frac{4\pi}{k^2}\sum_{l=0}^{\infty}(2l+1)
\sin^2\delta_l=4\pi\sum_{l=0}^{\infty}(2l+1)|a_l|^2\end{equation}which, for $\theta=0$, reduces to the \it
optical theorem\footnote{Introduced by Melvin Lax in (Lax 1950a,b).} $\sigma_{tot}(k)=(4\pi/k)\Im f(0)$. \rm
From (17),\marginpar{\tiny\it Breit-Wigner formula, Born approximation} it follows that, if the phase
$\delta_l=\delta_l(k)$ assumes, for a certain energy value\footnote{Indeed $E=\hbar^2k^2/2m$.} of $k$, the value
of $\pi$, then $a_l(k)=0$ and therefore the $l$-th partial wave does not concur to the scattering process (\it
Ramsauer-Townsend effect\rm), whilst if the phase $\delta_l$ assumes, for a certain energy value of $k$, the
value of $\pi/2$, then the contribution of the $l$-th partial wave will be maximally exalted in the scattering
process (\it resonance scattering\rm). In the latter case, the scattering contribution by the $l$-th partial
wave has the highest influence only when the phase $\delta_l$ changes in a neighborhood of the energy $E_0$
corresponding to that value of $k$ for which $\delta_l$ assumes the value of $\pi/2$, in this case being
possible to say that there exists an $l$-th partial wave resonance for $E=E_0$. In such a neighborhood, say
$\mathcal{I}(E_0)$, an approximate expression for the $l$-th amplitude $a_l$ was provided by G. Breit and E.P.
Wigner in 1946 (\it Breit-Wigner formula\rm), and given by
\begin{equation}a_l\approx -\frac{1}{k}\frac{\Gamma/2}{E-E_0+i(\Gamma/2)}\end{equation}where
$\Gamma=2/(d\delta_l(E)/dE)_{E=E_0}$. It provides a contribution to the total scattering cross-section given by
\begin{equation}\sigma_l(E)\equiv 4\pi(2l+1)|a_l|^2\approx\frac{2\pi\hbar^2}{m}(2l+1)\frac{1}{E}
\frac{\Gamma^2/4}{(E-E_0)^2+\Gamma^2/4}\end{equation}whose graph in function of $E$ shows a maximum value in the
neighborhood $\mathcal{I}(E_0)$ having a width approximately equal to $\Gamma$. The appearance of an $l$-th wave
function resonance peak physically corresponds to the production of a \it metastable \rm(or \it resonant\rm) \it
state\rm, localized in the interaction domain and with an equal angular momentum $l$, in which a particle may
stay for a sufficiently long time estimated by the mean lifetime given by $\tau=\hbar/\Gamma$, in such a manner
to exalt the corresponding scattering cross-section, this being the key essence of a scattering resonance
phenomenon. The \it Born approximation \rm for phases and scattering amplitudes respectively are
\begin{equation}\delta_l=-k\int_0^{\infty}U(r)[rj_l(kr)]^2dr, \ \ \ f(\theta)=-\int_0^{\infty}U(r)
\frac{\sin qr}{qr}r^2dr.\end{equation}

We are particularly interested in the complex case. When the potential has real values, then the scattering
phases are also real, but when the potential has complex values, then, in general, the scattering phases too are
complex. In the case of central potentials, as regard the scattering amplitude $f(\theta)$ given by (16), we
consider the following formula
\begin{equation}\begin{split}f(\theta)=&\frac{1}{2ik}\sum_{l=0}^{\infty}(2l+1)(e^{2i\delta_l}-1)P_l(\cos\theta)=\\
=&\frac{1}{2ik}\sum_{l=0}^{\infty}(2l+1)(S_l-1)P_l(\cos\theta),\end{split}\end{equation}having put $S_l\equiv
e^{2i\delta_l}$. As regard the elastic differential cross-section, we have $\sigma_{el}(\theta)=|f(\theta)|^2$,
while, taking into account (19), for the elastic total cross-section, we have
\begin{equation}\begin{split}\sigma_{el}^{tot}(k)=&\frac{\pi}{k^2}\sum_{l=0}^{\infty}(2l+1)|S_l-1|^2=\\
=&\frac{\pi}{k^2}\sum_{l=0}^{\infty}(2l+1)[1-2e^{-2\Im\delta_l}\cos(e\Re\delta_l)+e^{-4\Im\delta_l}].
\end{split}\end{equation}Instead, the inelastic differential scattering cross-section is given by
\begin{equation}\sigma_{in}(k)=\frac{\pi}{k^2}\sum_{l=0}^{\infty}(2l+1)[1-|S_l|^2]=
\frac{\pi}{k^2}\sum_{l=0}^{\infty}(2l+1)[1-e^{-4\Im\delta_l}].\end{equation}Therefore, in those scattering
processes in which are present many alternative phenomena, like elastic and inelastic collisions, it is needed
to consider the following expression for the total scattering cross-section\begin{equation}\begin{split}
\sigma_{tot}(k)=&\sigma_{el}^{tot}(k)+\sigma_{in}(k)=\frac{2\pi}{k^2}\sum_{l=0}^{\infty}(2l+1)[1-\Re S_l]=\\
=&\frac{2\pi}{k^2}\sum_{l=0}^{\infty}(2l+1)[1-2e^{-2\Im\delta_l}\cos(2\Re\delta_l)].
\end{split}\end{equation}

The temporal evolution\marginpar{\tiny\it S matrix, optical theorem} $t_0\rightarrow t$ of a quantum system is
defined through a unitary operator $S(t,t_0)$ which acts on an initial state $|s(t_0)\rangle$ producing the
final state $|s(t)\rangle=S(t,t_0)|s(t_0)\rangle$. In a scattering process, it is not necessary to know
$S(t_0,t)$ for arbitrary values of the time, but it is often enough to consider time intervals for $t_0\ll 0$
and $t\gg 0$ when one supposes that the interaction is efficient only into a definite time interval centered
around $t=0$; in such a case, we may define a unitary operator $S$, called \it $S$ matrix \rm(or \it scattering
matrix\rm) and introduced by J.A. Wheeler\footnote{Following (Mackey 1978, Section 21) and (Chew 1962, 1966,
Chapter 1), the idea of using the $S$ matrix (more properly, the $S$ operator) to describe the ''scattering'' of
particles by one another in quantum mechanics, was introduced by J.A. Wheeler in 1937 and again independently by
W. Heisenberg in 1943 who however lost interest in such a tool because of formal problems related just to its
analytic continuation that was required to give to the $S$ matrix dynamical content. Later, C. M\o ller
published two seminal papers, in 1945 and 1946, in which the mathematics underlying Heisenberg ideas were
developed, so that sometimes $S$ matrix is also called \it M\o ller wave matrix. \rm For a deeper historical
analysis, see for instance (Chew 1962, 1966), (Cushing 1986) and (Mussardo 2007, Part IV, Chapter 17, Appendix
A).} in 1937, as a limit of unitary operators in the following fashion
\begin{equation}S=\lim_{t_0\rightarrow-\infty\atop t\rightarrow\infty}S(t_0,t)\end{equation}where the asymptotic
states $t_0=-\infty$ and $t=\infty$ refer to ideal moments in which the dynamical system is supposed to be into
a non-interacting state. The $S$ matrix (together the so-called \it $T$ matrix, \rm defined by
$S=\mathbb{I}+iT$) is a formal tool which was mainly set up to study relativistic interactions of elementary
subnuclear particles whose related forces were so little known in the form and nature to entail a quantum
description through the Schr\"{o}dinger and Dirac equations almost unusable. Instead, by means of such a tool,
the scattering amplitudes could be deduced from general physical and mathematical principles no matter by the
knowledge of the involved forces. Furthermore, the proliferation of the impetuous phenomenology of 1950s
high-energy physics led to a so great amount of new and unexpected phenomena that the old quantum field theory
turned out to be unable to provide a right theoretical basis to them, to make necessary appealing to general
formal principles, above all unitarity and analyticity, to theoretically explain these new results. Every
element of the $S$ matrix, that is to say $S_{fi}=\langle f|S|i\rangle$, is the probability amplitude to observe
a particular final state $|f\rangle$ starting from a given initial state $|i\rangle$, that is to say, the
probability amplitude associated to the dynamical process $|i\rangle\rightarrow|f\rangle$. Other basic formal
properties of the $S$ matrix may be deduced from general physical principles, like relativistic invariants. For
instance, as regard collision processes involving four scalar particles of the type $1+2\rightarrow 3+4$, the
possible relativistic invariants are $s=(p_1+p_2)^2$, $t=(p_1-p_2)^2$ and $u=(p_1-p_2)^2$, which satisfy the
following functional dependence relation $s+t+u=\sum_{i=1}^4m_i^2$ where $m_i^2=p_i^2=p_{0i}^2-\vec{p}_i^2$.
Moreover, another fundamental property of $S$ matrix is the so-called \it crossing \rm which relates the
scattering amplitude of the initial process $1+2\rightarrow 3+4$ with those processes obtained by this replacing
one incoming particle with momentum $p_i$ of the former process with its antiparticle having four-momentum
$-p_i$, and vice versa. For example, in terms of $T$ matrix, if $T(s,t,u)$ is the scattering amplitude of the
process $1+2\rightarrow 3+4$, then it is also the scattering amplitude of the process
$1+\bar{3}\rightarrow\bar{2}+4$ or of the process $1+\bar{4}\rightarrow\bar{2}+3$. Therefore, information about
different collision processes may be suitably put into analytical relationship between them: for instance, one
of these is given by that providing the presence of poles in correspondence to possible bound states or
resonances. The scattering amplitude cannot have only poles because them generate only a real amplitude whilst
the unitarity condition implies the existence of an imaginary part, say $\Im T(s,u,t)$, proportional to the
total scattering cross-section, and that, in turn, contains the various cuts of the corresponding Riemann
surface. Therefore, crossing and unitarity link together analytical properties of different collision processes,
that is to say, different elements of the $S$ (or $T$) matrix, a notable fact, this, which was overstated in the
1960s till to think that such a formal scheme\footnote{Upon which relies the so-called \it bootstrap hypothesis
\rm according to which, roughly speaking, from a given set of bound states - named ''particles'' - describing
physical states and forces through a well-defined dynamics formulated in such a manner that other states may be
self-consistently generated without make reference to initial fundamental states.} could determine all the
scattering amplitudes through well-determined spectral representations, amongst which certain dispersion
relations\footnote{Following (Cini 1977), the dispersion relations had a preeminent role in high-energy physics,
becoming the prevalent paradigm for strong interactions, within which framework a particular analytical property
of the scattering amplitude related to non-relativistic Schr\"{o}dinger equation was extended, by analogy, to
the relativistic context just by Regge's work. See (Cini 1977) for a deep historical account and basic
philosophy of science considerations related to dispersion relations.}, involving the various singularities of
the analytic $S$ (or $T$) matrix.

Setting\begin{equation}|s_i\rangle=\lim_{t\rightarrow-\infty}|s(t)\rangle,\ \ \ |s_f\rangle=
\lim_{t\rightarrow\infty}|s(t)\rangle,\end{equation}we have
\begin{equation}|s_f\rangle=S|s_i\rangle,\end{equation}so that into the matrix $S$ are included all the
information on the scattering process by which the initial state $|s_i\rangle$ temporally evolves to the final
state $|s_f\rangle$ due to the existence of an interaction between the particles constituting the given
dynamical system involved into the scattering process. Now, if $|s_i\rangle=|a\rangle$, then the final state has
the following expansion into a set of complete orthonormal states $|b\rangle$
\begin{equation}|s_f\rangle=\int C_{ab}|b\rangle=\int\langle b|s_f\rangle|b\rangle=\int\langle b|S|a\rangle|b\rangle
=\int S_{ab}|b\rangle,\end{equation}where the index $a$ refers to the state evolution
$|a\rangle\rightarrow|b\rangle$, while $S_{ab}=\langle b|S|a\rangle$, thanks to the unitarity of $S$, provides
the probability amplitude, given by $W_{ab}=|S_{ab}|^2$, with which the dynamical system may run along
scattering process from $|a\rangle$ to $|b\rangle$. In general, the final state $|b\rangle$ is, roughly
speaking, characterized by a continuous variability in such a manner that it is more correct to formally
consider a differential probability defined as $dW_{ab}=|\langle b|S|a\rangle|^2db$ which, therefore, is the
transition probability from the initial state $|a\rangle$ into one of the final states $|b\rangle$ having
quantum numbers comprised between $b$ and $b+db$. In the time-dependent interaction representation, the $S$
matrix has the following representation
\begin{equation}S=Te^{\displaystyle-\frac{i}{\hbar}\int_{-\infty}^{\infty}H_I'(t)dt}\end{equation}said
to be the \it Dyson's formula\rm, $T$ being the time decreasing operator called \it chronological product \rm
and $H'$ the perturbative part of the hamiltonian operator. At the first order perturbative effects, this
formula reduces to the following \it Born approximation \rm for $S$\begin{equation}S\approx
S_B=\mathbb{I}-\frac{i}{\hbar}\int_{-\infty}^{\infty}H_I'(t)dt.\end{equation}In a scattering process, over each
energy surface $E$, we have the following conservation energy relation
\begin{equation}\langle b|S|a\rangle=\langle b|S^E|a\rangle\delta(E_b-E_a)\equiv S_{ba}^E\delta(E_b-E_a)\end{equation}
where $S_{ba}^E\equiv\langle b|S^E|a\rangle$ is the so-called $S$ matrix element over the energy surface $E$.
Because of the continuous variability of the final states $b$, we have
\begin{equation}dW_{ba}=\frac{1}{2\pi\hbar}|\langle b|S^E|a\rangle|^2\frac{db}{dE_b}
\overset{Born\atop approx.}{\longrightarrow}\frac{2\pi}{\hbar}|\langle
b|H_I'|a\rangle|^2\rho_f(b)\end{equation}which is said to be the \it first Fermi golden rule\rm, where
\begin{equation}\rho_f(b)=db/dE_b\end{equation}is the density of the final states. With greater precision, the two
above mentioned quantum states, involved in the scattering process and considered satisfying the right
normalization conditions, should be denoted as follows $|a,E_a,\vec{n}_a\rangle$ and $|b, E_b,
\vec{n}_b\rangle$, where $a,b$ refer to the nature of the system and to the related characterizing quantum
numbers, while $\vec{n}_a=\vec{p}_a/p_a=\vec{k}_a/k_a$ and $\vec{n}_b=\vec{p}_b/p_b=\vec{k}_b/k_b$ identify the
related momentums whose directions are respectively identifiable by means of the related solid angles
$\Omega_a=(\theta_a,\varphi_a)$ and $\Omega_b=(\theta_b,\varphi_b)$. In particular, we have $db=dE_bd\Omega_b$,
so that, in the hypothesis $E=E_b=E_a$, the first right hand side of (34) becomes
\begin{equation}dW_{ba}=\frac{1}{2\pi\hbar}|\langle
b,E,\vec{n}_b|S^E|a,E,\vec{n}_a\rangle|^2d\Omega_b\end{equation}which provides the transition's probability per
unit of time, from the initial state $|a\rangle$ to the final state $|b\rangle$ with wave vector having
direction $\vec{n}_b$ comprised into the elementary solid angle $d\Omega_b$. Making use of the probability
current density associated to a quantum particle, it is possible to prove that the differential scattering
cross-section for the elastic and inelastic scattering processes $a\rightarrow b$ is given by
\begin{equation}\sigma_{ba}(\Omega_b)d\Omega_b=\frac{dW_{ba}}{|\vec{j}_a|}=\frac{4\pi^2}{k_a^2}
|\langle b,E,\vec{n}_b|S^E-\mathbb{I}|a,E,\vec{n}_a\rangle|^2d\Omega_b,\end{equation}where $\vec{j}_a$ is the
probability current density associated with the particles of the incident beam. For elastic scattering
processes, since $a=b$, we may write
\begin{equation}\sigma(\Omega)\equiv\sigma(\theta,\varphi)=\frac{4\pi^2}{k^2}|\langle\vec{n}'|S^E-\mathbb{I}|\vec{n}
\rangle|^2\end{equation}where $\Omega\equiv(\theta,\varphi)$ identifies the direction $\vec{n}'$ of the
scattered wave vector with respect to the direction of the incident wave vector $\vec{n}$. From (6) and (38), it
follows that
\begin{equation}f(\theta,\vartheta)=\frac{2\pi}{k}e^{i\alpha}\langle\vec{n}'|S^E-\mathbb{I}|\vec{n}\rangle\end{equation}
where $e^{i\alpha}$ is a suitable phase factor.

In a central potential problem, the angular momentum $\vec{L}$ is a constant of motion so that, said
$|lm\rangle$ an orthonormal basis of eigenstates of the angular momentum operators $L^2$ and $L_z$, the $S$
matrix elements are diagonal and we may set\begin{equation}\langle
b,l'm'|S^E|a,lm\rangle=\delta_{ll'}\delta_{mm'}S_{ba}^l\end{equation}where the reduced $S$ matrix elements
$S_{ba}^l(K)$ does not depend on $m$, seen the arbitrariness of the direction $z$ with respect to which compute
$L_z$, but only on the two states $a,b$ and on the energy $E$ via the modulus $k$ of the relative wave vector.
The elements of the operator $S^E-\mathbb{I}$ upon the energy surface $E$, given by
\begin{equation}\langle b,\vec{n}_b|S^E-\mathbb{I}|a,\vec{n}_a\rangle,\end{equation}can be estimated as follows
\begin{equation}\langle b,\vec{n}_b|S^E-\mathbb{I}|a,\vec{n}_a\rangle=\frac{1}{4\pi}\sum_{l=0}^{\infty}
(2l+1)(S_{ba}-\delta_{ba})P_l(\cos\theta)\end{equation}where $\theta$ is the scattering angle between
$\vec{n}_a$ and $\vec{n}_b$, such a formula providing the Legendre polynomial expansion of the matric elements
(39), with coefficients which depend on the reduced $S$ matrix elements $S_{ba}^l$. For elastic scattering
processes, $a=b$ so that, setting $S_{aa}^l\equiv S_l$, we have\begin{equation}\langle
\vec{n}'|S^E-\mathbb{I}|\vec{n}\rangle=\frac{1}{4\pi}\sum_{l=0}^{\infty}
(2l+1)(S_l-1)P_l(\cos\theta).\end{equation}But, from (39) for central symmetry problems, we have
\begin{equation}f(\theta)=\frac{2\pi}{k}e^{i\alpha}\langle
\vec{n}'|S^E-\mathbb{I}|\vec{n}\rangle\end{equation}with $\alpha\in\mathbb{R}$ an undefined parameter, so
that\begin{equation}f(\theta)=\frac{e^{i\alpha}}{2k}\sum_{l=0}^{\infty}
(2l+1)(S_{l}-1)P_l(\cos\theta)\end{equation}which, compared with the following Legendre polynomial expansion
\begin{equation}f(\theta)=\frac{1}{2ik}\sum_{l=0}^{\infty}(2l+1)(e^{2i\delta_l}-1)P_l(\cos\theta),\end{equation}implies
$e^{i\alpha}=-i$ and $S_l=e^{2i\delta_l}$, that is to say, the reduced $S$ matrix elements are closely related
to the corresponding phase displacements $\delta_l$. Therefore, (39) reduces to
\begin{equation}f(\theta,\varphi)=-\frac{2\pi
i}{k}\langle\vec{n}'|S^E-\mathbb{I}|\vec{n}\rangle\end{equation}which holds too for all the elastic scattering
processes, comprised the non-central ones, being $\theta$ and $\varphi$ the polar angles of $\vec{n}'$ with
respect to the incident direction $\vec{n}$. In the case of central potential scattering, using (42) in (37), we
obtain the following expression for the differential scattering cross-section
\begin{equation}\sigma_{ba}(\Omega_b)=\frac{1}{4k^2}|\sum_{l=0}^{\infty}(2l+1)(S_{ba}^l-\delta_{ba})
P_l(\cos\theta)|^2,\end{equation}whilst for the total scattering cross-section, we have
\begin{equation}\sigma_{ba}^{tot}(k)=\frac{\pi}{k^2}\sum_{l=0}^{\infty}(2l+1)
|S_{ba}^l-\delta_{ba}|^2,\end{equation}both these latter formulas being valid for every scattering process
$a\rightarrow b$. For elastic scattering processes ($a=b$), they respectively reduce to the following
\begin{equation}\sigma_{ba}^{el}(\theta)=\frac{1}{4k^2}|\sum_{l=0}^{\infty}(2l+1)(S_l-\mathbb{I})
P_l(\cos\theta)|^2\end{equation}and\begin{equation}\sigma^{tot}_{el}(k; a\rightarrow
b)=\frac{\pi}{k^2}\sum_{l=0}^{\infty}(2l+1) |S_l-\mathbb{I}|^2,\end{equation}whereas, for inelastic scattering
processes ($a\neq b$), we respectively have
\begin{equation}\sigma_{ba}^{inel}(\theta)=\frac{1}{4k^2}|\sum_{l=0}^{\infty}(2l+1)
S_{ba}^lP_l(\cos\theta)|^2\end{equation}and\begin{equation}\sigma^{tot}_{inel}(k;a\rightarrow b)=
\frac{\pi}{k^2}\sum_{l=0}^{\infty}(2l+1)|S_{ba}^l|^2.\end{equation}Often, a sum over all the possible scattering
channels $a\rightarrow b$ is considered as follows
\begin{equation}\begin{split}\sigma^{tot}_{inel}(k)=&\sum_{a\neq
b}\sigma^{tot}_{ba}(k)=\frac{\pi}{k^2}\sum_{l=0}^{\infty}(2l+1)\sum_{a\neq b}|S_{ba}^l|^2=\\
=&\frac{\pi}{k^2}\sum_{l=0}^{\infty}(2l+1)(1-|S_l|^2).\end{split}\end{equation}Furthermore, in both elastic and
inelastic scattering process case, we have $S_l=e^{2i\delta_l}$ with $|S_l|\leq 1$, so that
$|S_l|=|e^{2i\delta_l}|=e^{-2\Im\delta_l}$ and $\Im\delta_l\geq 0$. As regard the total scattering
cross-section, we have
\begin{equation}\sigma_{tot}(k)=\sigma_{el}^{tot}(k)+\sigma_{inel}^{tot}(k)=\frac{2\pi}{k^2}\sum_{l=0}^{\infty}(2l+1)
(1-\Re S_l).\end{equation}On the other hand, taking into account (46), we have
\begin{equation}\Im f(0)=\frac{1}{2k}\sum_{l=0}^{\infty}(2l+1)(1-\Re S_l),\end{equation}that, compared with (55),
gives rise to the \it optical theorem \rm in its most general form
\begin{equation}\Im f(0)=\frac{k}{4\pi}\sigma_{tot}(k).\end{equation}In analogy with (47), which links together
the elastic scattering amplitude with the elements of the $S$ matrix, we may define a \it generalized scattering
amplitude \rm as follows\begin{equation}f_{ba}(\vec{n}_b,\vec{n}_a)\equiv f_{ba}(\Omega_b)=-\frac{2\pi
i}{k}\langle b\vec{n}_b|S^E-\mathbb{I}|a\vec{n}_a\rangle\end{equation}where $\Omega_{b}=\Omega\equiv
(\theta,\varphi)$ is the solid angle which identifies the direction of $\vec{n}_b$ with respect to $\vec{n}_a$,
this last chose parallel to the direction of the polar axis in such a manner that the differential scattering
cross-section for the channel $a\rightarrow b$, is given by $\sigma_{ba}(\Omega_b)=|f_{ba}(\Omega_b)|^2$. In the
case of multichannel processes, for simplicity's sake we consider a three channel process whose directions are
$\vec{n}, \vec{n}'$ and $\vec{n}''$, by which we have
\begin{equation}f(\vec{n}',\vec{n})-f^*(\vec{n},\vec{n}')=\frac{ik}{2\pi}\int f^*(\vec{n}'',\vec{n}')
f(\vec{n}'',\vec{n})d\Omega''\end{equation}that is the general condition to which undergoes the scattering
amplitude for purely elastic processes, due to unitarity of $S$, and that, for central collisions, reduces to
the optical theorem. Indeed, in such a case, $f(\vec{n}', \vec{n})$ depends on $\cos\theta=\vec{n}\cdot\vec{n}'$
only, and since we have too $f(\vec{n},\vec{n}')=f(\vec{n}',\vec{n})$, it follows that (58) reduces
to\begin{equation}\Im f(\vec{n},\vec{n}')=\frac{k}{4\pi}\int f^*(\vec{n}'',\vec{n}')
f(\vec{n}'',\vec{n})d\Omega''\end{equation}that is an extension of the optical theorem because, for
$\vec{n}=\vec{n}'$, we have $\theta=0$, so that (60) reduces further to\begin{equation}\Im
f(\vec{n},\vec{n})=\frac{k}{4\pi}\int
|f(\vec{n}'',\vec{n})|^2d\Omega''=\frac{k}{4\pi}\sigma_{tot}\end{equation}that is to say, the optical theorem
(57). It is also interesting to notice that (60) allows, at least in principle, to rebuild up the scattering
amplitude when it is known its modulus, that is to say, it is known the differential cross-section, excepts the
indeterminacy arising from the invariance of (60) with respect to the change $f(\theta)\rightarrow
-f^*(\theta)$.

Many fundamental\marginpar{\tiny\it Regge poles, Jost solutions and functions} results of scattering theory may
be deduced without make reference to a specific form of the related interaction, only through the basic
properties of the $S$ matrix like, for instance, its unitarity from which we have information about the
scattering amplitude. Further properties may be deduced from other analyticity properties of the elements of the
$S$ matrix, which are able to describe a collision process. In what follows, we shall refer to elastic
collisions in a central field, where the $S$ matrix elements are given by the functions $S_l\equiv S_l(k)$ -
that is to say, $S$ matrix elements computed over a surface of energy\footnote{Recall that $E=\hbar^2k^2/2\mu$.}
$k$ with values related to states of definite angular momentum $l$ - depending on the scattering phases
through\begin{equation}S_l(k)=\exp(2i\delta_l(k)).\end{equation}For instance, for an elastic scattering process
in a central field, if one considers (16) wrote in the form
\begin{equation}f(k,\theta)=\sum_{l=0}^{\infty}(2l+1)a_l(k)P_l(\cos\theta),\end{equation}where the partial amplitudes
$a_l(k)$ are given by (17) and written as follows
\begin{equation}a_l(k)=\frac{1}{2ik}(S_l(k)-1),\end{equation}then it follows that properties of $f(k,\theta)$
are consequence of the properties either of $S_l(k)$ and of the Legendre polynomials $P(\cos\theta)$. The latter
formulas (63) and (64) resolve, d'après H. Faxen and J. Holtsmark works of 1927, the problem how to express the
scattering amplitude by means of the scattering phases. Moreover, from the orthogonality properties of these
latter, an inversion of (63) gives rise to
\begin{equation}a_l(k)=\frac{1}{2}\int_{-1}^{1}f(k,\theta)P_l(\cos\theta)d\cos\theta\end{equation}hence, from
(64), it also follows
\begin{equation}S_l(k)=1+ik\int_{-1}^1f(k,\theta)P_l(\cos\theta)d\cos\theta,\end{equation}so that there exists a
certain equivalence between the informations given by the scattering amplitude $f(k,\theta)$ and those given by
the set of functions $S_l(k)$. Therefore, it is of fundamental importance to study the analytical properties of
the functions $S_l(k)$ considered as function either of $l$ and $k$, though only for $k\in\mathbb{R}^+$ and
$l\in\mathbb{N}_0$ they have a direct physical meaning for scattering processes. Nevertheless, as we will see
later, the consideration of these functions also for complex values of both these two variables, gives rise to
interesting and unexpected results and interpretations. For instance, the knowledge of the properties of
analyticity in the variable $k$ of the functions $S_l(k)$, allows, inter alia, to show that the scattering
amplitude verifies certain dispersion relations in the variable $k$, whereas the knowledge of the properties of
analyticity in the variable $l$ of the functions $S_l(k)$, provides a representation for the scattering
amplitude which is useful to study the asymptotic properties for great transferred impulses, so giving rise to
the so-called \it Regge poles. \rm Therefore, for pursuing this, it is preliminarily need to define $S_l(k)$ for
arbitrary values of $l$ and $k$, and this may be done, in potential theory\footnote{Hence, within the
non-relativistic quantum mechanics framework where an interaction's potential is definable, so that the
dynamical equation governing the collision process is the Schr\"{o}dinger equation.}, through the following
reduced radial Schr\"{o}dinger equation
\begin{equation}u''_l+[k^2-\frac{l(l+1)}{r^2}-U(r)]u_l=0\end{equation}which, mathematically, has always a meaning
for every complex value of $k$ and $l$. The relation (64) has been obtained by taking into consideration a wave
function with an asymptotic behavior given by (5) and developed in partial waves as follows
\begin{equation}\psi(\vec{r})=\sum_{l=0}^{\infty}i^l(2l+1)\frac{\chi_l(k,r)}{r}P_l(\cos\theta)\end{equation}where
$\chi_l(k,r)$ are the physical solutions to (67) which are zero in the origin of the given reference frame, and
having the following asymptotic behavior
\begin{equation}\chi_l(k,r)\underset{r\rightarrow\infty}{\sim}\frac{e^{i\delta_l(k)}}{k}
\sin[kr-l\frac{\pi}{2}+\delta_l(k)]\end{equation}which, in absence of any interaction, reduces the following
free solutions\begin{equation}\chi^0_l(k,r)=rj_l(k,r)\end{equation}where $j_l(k,r)$ are the Bessel spherical
functions. The formula (69) implicitly defines the scattering phase $\delta_l(k)$, hence $S_l(k)$ via (62),
through $\chi_l(k,r)$ for each $l,k$. Thus, the analyticity properties of the functions $S_l(k)$ will be known
once we know those of the functions $\chi_l(k,r)$ as solutions to (67). For simplicity's sake, we shall consider
simple solutions to (67) as, for instance, those obtained under certain boundary (in the origin or at infinity)
conditions which are very simple and which are dependent only on one of the two variable $k$ and $l$, from
which, therefore, to accordingly derive fundamental analyticity properties of $S_l(k)$ in the variable $k$ (with
$l$ fixed at a given physical value) or in the variable $l$ (with $k$ fixed and belonging to $\mathbb{R}^+$). If
we set $\lambda=l+1/2$, then (67) reduce to the following more symmetric form
\begin{equation}u''+[k^2-\frac{\lambda^2-1/4}{r^2}-U(r)]u=0\end{equation}whose solutions are of the form
$u=u(\lambda,k,r)$, or else $u=u_l(k,r)$ when we wish to refer to physical values of $l$. We search for
solutions to (71) through suitable hypotheses on the interaction potential $U(r)$ which, nevertheless, allow us
to have an enough degree of generality to leave aside from the particular type of potential so chosen. Such
hypotheses regard the behavior of the potential either in the origin of the reference frame and at infinity. To
be precise, we consider short-range potentials $U(r)$ such that
\begin{equation}\begin{split}\int_0^ar|U(r)|dr=&M(a)<\infty\quad\forall a\in\mathbb{R}^+,\\\\
\int_b^{\infty}|U(r)|dr=&N(b)<\infty\quad\forall b\in\mathbb{R}^+.\end{split}\end{equation}The hypothesis
$(72)_1$ implies that the point $r=0$ is a singular point of Fuchsian type for the equation (71), with
characteristic exponents $\lambda+1/2\equiv l+1$ and $-\lambda+1/2\equiv -l$, so that surely there exist two
solutions to (71) which behave like to $r^{\pm\lambda+1/2}$ as $r\rightarrow 0$. Instead, the hypothesis
$(72)_2$ implies that (71) has solutions which have an asymptotic behavior similar to the one of the solutions
to the equation $u''+k^2u=0$, so that surely there exist solutions to (71) which behave like to $\exp(\pm
i\vec{k}\cdot\vec{r})$ as $r\rightarrow\infty$. Therefore, through such asymptotic behaviors, namely
$r^{\pm\lambda+1/2}$ as $r\rightarrow 0$ and $e^{\pm i\vec{k}\cdot\vec{r}}$ as $r\rightarrow\infty$, we are able
to find solutions to (71).

The solution $u=\phi(\lambda, k,r)\underset{r\rightarrow 0}{\sim}r^{\lambda+1/2}$, with the boundary condition
\begin{equation}\lim_{r\rightarrow 0}r^{-\lambda-1/2}\phi(\lambda,k,r)=1,\end{equation}is said to be the \it regular
solution \rm to (71), and, as such, it turns out to be given by
$\phi(\lambda,k,r)\equiv\phi_l(k,r)=C_l(k)\chi_l(k,r)$. Likewise, the function $\phi(\lambda,k,r)$, in absence
of interaction, must reduce to the free solution $\phi^0(k,r)$, that is to say, to the solution to the free
equation\begin{equation}u''+\Big(k^2-\frac{\lambda^2-1/4}{r^2}\Big)u=0\end{equation}which has the same
asymptotic behavior of $\phi(\lambda, k, r)$ in $r=0$. For $l\in\mathbb{N}_0$, we have
\begin{equation}\phi^0(k,r)=\frac{(2l+1)!!}{k^l}rj_l(k,r)\end{equation}where $j_l(k,r)$ are the spherical Bessel
functions. Therefore, to may carry out an extension to every value of $l$, we should consider the functional
generalization of the semi-factorial, hence the Euler gamma function, so that we have
\begin{equation}\phi^0(\lambda,k,r)=2^{\lambda}k^{-\lambda}\Gamma(1+\lambda)r^{1/2}J_{\lambda}(kr)\end{equation}where
$J_{\lambda}(kr)=J_{l+1/2}(kr)$ are the spherical Bessel functions of the first kind, defined for arbitrary
values of $l$. To find analytical properties of $\phi(\lambda,k,r)$ as a solution to (71), we may construct a
suitable integral equation of the Volterra type to which such a $\phi$ must satisfy under the boundary
conditions (72), and this, in turn, may be accomplished by means of an auxiliary equation that takes into
account the same boundary conditions (72), equivalent to\begin{equation}\phi\underset{r\rightarrow
0}{\sim}r^{\lambda+1/2},\qquad \phi'\underset{r\rightarrow 0}{\sim}(\lambda+1/2)r^{\lambda-1/2},\end{equation}
and that has known solutions having the same behavior, in the origin of the reference frame, of the searched
solutions to (71), that is to say, having a behavior as $r^{\pm\lambda+1/2}$. As a possible auxiliary equation
to which $\phi$ ought to be satisfied as $r\rightarrow 0$, we choose the following one
\begin{equation}u''-\frac{\lambda^2-1/4}{r^2}u=0,\end{equation}which has the following pair of fundamental
solutions\begin{equation}u_{+}=r^{\lambda+1/2},\qquad u_-=r^{-\lambda+1/2}.\end{equation}It is possible to prove
that such an integral equation is as follows
\begin{equation}\begin{split}&\phi(\lambda,k,r)=u_+(r)+\\&+\frac{1}{2\lambda}\int_0^r[k^2-U(x)]
[u_+(x)u_-(r)-u_-(x)u_+(r)]\phi(\lambda,k,x)dx,\end{split}\end{equation}so that, the integral equation of the
Volterra type to which the regular solution $\phi(\lambda,k,r)$ must satisfy is
\begin{equation}\begin{split}\phi(\lambda,k,r)&=r^{\lambda+1/2}+\\&+\frac{1}{2\lambda}\int_0^r[k^2-U(x)]
\sqrt{rx}\Big\{\Big(\frac{x}{r}\Big)^{\lambda}-\Big(\frac{r}{x}\Big)^{\lambda}\Big\}
\phi(\lambda,k,x)dx\end{split}\end{equation}whose formal solution is
\begin{equation}\phi(\lambda,k,r)=\sum_{n=0}^{\infty}\phi_n(r)\end{equation}with $\phi_0(r)=r^{\lambda+1/2}$ and
\begin{equation}\phi_n(r)=\frac{1}{2\lambda}\int_0^r[k^2-U(x)]
\sqrt{rx}\Big\{\Big(\frac{x}{r}\Big)^{\lambda}-\Big(\frac{r}{x}\Big)^{\lambda}\Big\}
\phi_{n-1}(x)dx,\quad\forall n\in\mathbb{N},\end{equation}which provides too an effective solution to (80) when
it is uniformly convergent. To study the related convergence, we distinguish between the two cases
$\Re\lambda\geq 0$ and $\Re\lambda<0$. Putting $\lambda=\mu+i\sigma$, in the case $\Re\lambda=\mu\geq 0$, which
comprises the physical case of angular momentum values (which are semi-integers), it is possible to prove that
\begin{equation}|\phi_n(r)|\leq\frac{r^{\mu+1/2}}{|\lambda|^n}\frac{[P(r)]^n}{n!},\quad\forall
n\in\mathbb{N},\end{equation}with $\displaystyle P(r)\doteq\int_0^rx|k^2-U(x)|dx\leq (1/2)|k^2|r^2+M(r)$, so
that we have
\begin{equation}|\phi(\lambda,k,r)|\leq\sum_{n=0}^{\infty}|\phi_n(r)|\leq
r^{\mu+1/2}\exp(P(r)/|\lambda|),\end{equation}a condition which guarantees convergence. Now, a celebrated
theorem, due to H.J. Poincaré, which roughly states that a solution to a differential equation, like (71), whose
coefficients are entire functions of a certain parameter, is also an entire function of the same parameter when
defined through boundary conditions which are independent from this parameter, may be applied to (81). Indeed,
(81) is, in every finite domain, a uniformly convergent series of polynomial in the variable $k^2$, so that it
is an entire function of $k^2$ as well, in agreement with the just above mentioned Poincaré theorem. Moreover,
it is possible to prove that, at least in the region $\Re\lambda>0$, each term of (81) is an analytic function
in the variable $\lambda$, so that also $\phi(\lambda,k,r)$ is analytic there. The analytic continuation of
$\phi(\lambda,k,r)$ for negative values of $\Re\lambda$ entails more detailed information about the potential
$U(r)$; in particular, if one supposes that\begin{equation}U(r)\underset{r\rightarrow
0}{\sim}cr^{-2+\epsilon},\qquad\epsilon>0,\end{equation}then it is possible to prove that $\phi(\lambda,k,r)$ is
analytic in $\lambda$ in the region $\Re\lambda>-\epsilon/2$, so that there exists, at least, a region of the
$\lambda$ plane, i.e. the strip $|\Re\lambda|<\epsilon/2$, in which the functions $\phi(\lambda,k,r)$ and
$\phi(-\lambda,k,r)$ are contemporaneously defined and constitute a fundamental pair of solutions to (71) for
$\lambda\neq 0$. Therefore, for each $\lambda$ arbitrarily fixed, $\phi(\lambda,k,r)$ is an entire function of
$k^2$, at least for $\Re\lambda>0$, or else for $\Re\lambda>0$ if (86) holds, as well as it is an even function
of $k$, i.e. $\phi(\lambda,-k,r)=\phi(\lambda,k,r)$; furthermore, for each $k$, $\phi(\lambda,k,r)$ is an
analytic function of $\lambda$ at least for $\Re\lambda>0$ or else for $\Re\lambda>-\epsilon/2$ if (86) holds,
for each finite value of $k$ arbitrarily fixed. Instead, as concerns the behavior of $\phi(\lambda,k,r)$ as
$k\rightarrow\infty$, it is possible to prove that, for any direction by which $k\rightarrow\infty$,
$\phi(\lambda,k,r)$ reduces to the following free solution
\begin{equation}\phi_0(\lambda,k,r)=\Big(c_1\frac{e^{i\vec{k}\cdot\vec{r}}}{k^{\lambda+1/2}}+
c_2\frac{e^{-i\vec{k}\cdot\vec{r}}}{k^{\lambda+1/2}}\Big)(1+O(1/k))\end{equation}where $c_1,c_2$ are independent
from $k$. Finally, we notice that $\phi(\lambda,k,r)$ is real when $k$ and $r$ are real (under the obvious
hypothesis that $U(r)$ is also real) because of the boundary conditions (73), so that, due to the H.A. Schwarz
reflection principle, the following Hermiticity condition holds $\phi^*(\lambda,k,r)=\phi(\lambda^*,k^*,r)$
which, for physical values of $l=\lambda-1/2$, implies $\phi_l^*(k,r)=\phi_l(k^*,r)$.

We have seen that the $S$ matrix elements, as well as the scattering phases, are determined by the behavior of
the wave function as $r\rightarrow\infty$, so that a primary role is played by boundary conditions at infinity,
like (72); in particular, the condition $(72)_2$ says that $U(r)$ goes to zero faster than $1/r$ as
$r\rightarrow\infty$, and guarantees the existence of two independent solutions to (71), which, as
$r\rightarrow\infty$, behave as $\exp(\pm i\vec{k}\cdot\vec{r})$. Let $f(\lambda,k,r)$ be the exact solution to
(71) with the following boundary condition
\begin{equation}\lim_{r\rightarrow\infty}e^{i\vec{k}\cdot\vec{r}}f(\lambda,k,r)=1,\end{equation}so that
\begin{equation}f(\lambda,k,r)\underset{r\rightarrow\infty}{\sim}e^{-i\vec{k}\cdot\vec{r}},\end{equation}which
is said to be the \it Jost solution. \rm Moreover, due to the inversion symmetry invariance of (71) with respect
to the variable $k$, it follows that $f(\lambda,-k,r)$ too is a solution to (71), with
\begin{equation}f(\lambda,-k,r)\underset{r\rightarrow\infty}{\sim}e^{i\vec{k}\cdot\vec{r}}.\end{equation}Therefore,
for every $k\neq 0$, we shall refer to the pair of Jost solutions $f(\lambda,\pm k,r)$ as two independent
fundamental solutions to (71). The free Jost solution is given by
\begin{equation}f_0(\lambda,k,r)=\sqrt{\frac{\pi kr}{2}}e^{-i(\lambda+1/2)\pi/2}H_{\lambda}^{(2)}(kr)\end{equation}where
$H_{\lambda}^{(2)}$ is the second-kind Hankel function. Since (88) does not depend on $\lambda$, the above
mentioned Poincaré theorem states that, for each $k$ fixed, $f(\lambda,k,r)$, where is defined, is an entire
function of $\lambda^2$, hence everywhere analytic over the finite plane of $\lambda$. As an entire function of
$\lambda^2$, $f(\lambda,k,r)$ is even in the variable $\lambda$, that is to say,
$f(-\lambda,k,r)=f(\lambda,k,r)$. Indeed, if one proceeds as made above, the analytical properties of
$f(\lambda,k,r)$ may be deduced transforming the differential equation to which it satisfies, i.e.
\begin{equation}f''+\Big[k^2-\frac{\lambda^2-1/4}{r^2}\Big]f=0\end{equation}with the boundary condition (88), into
an appropriate integral equation of the Volterra type, through the following auxiliary equation
\begin{equation}u''+k^2u=0\end{equation}and its solutions\begin{equation}u_+=e^{i\vec{k}\cdot\vec{r}},\qquad
u_-=e^{-i\vec{k}\cdot\vec{r}}.\end{equation}Such an integral equation is
\begin{equation}\begin{split}f(\lambda,k,r)=& e^{-i\vec{k}\cdot\vec{r}}+\\+&\frac{1}{k}
\int_r^{\infty}\sin[k(x-r)]\Big[U(x)+\frac{\lambda^2-1/4}{x^2}\Big]f(\lambda,k,x)dx\end{split}\end{equation}from
which it follows that $f(\lambda,k,r)$, for each $k\neq 0$, is an entire function of $\lambda^2$ reasoning as
made above by means of a series expansion whose general $n$-th term is a polynomial of degree $n$ in the
variable $\lambda^2$, at least in the region $\Im k<0$. Likewise, it is possible to prove that, for any
direction by which $|k|\rightarrow\infty$, we have
\begin{equation}f(\lambda,k,r)\underset{|k|\rightarrow\infty}{\sim}f_0(\lambda,k,r).\end{equation}The
analyticity region in $k$ may be extended if, for instance, one considers potentials $U(r)$ such that
\begin{equation}\int_0^{\infty}rU(r)e^{\nu r}dr<\infty\qquad\mbox{for}\ \nu<m,\end{equation}where $m$ is an arbitrary
positive number, then it is possible to prove that $f(\lambda,k,r)$ is analytic in $k$ in the region $\Im
k<m/2$, except $k=0$, which is a polydromy point having kinematical nature because it is also shared by
$f_0(\lambda,k,r)$, besides to become a pole of order $l$ for physical angular momenta ($l\in\mathbb{N}_0$).
Taking into account this, it is usual to cut the complex plane\footnote{Following (Eden et al., 1966, Preface),
one of the most remarkable discoveries in elementary particle physics has been that of the existence of the
complex plane. From the early days of dispersion relations, the analytic approach to the subject has proved to
be one of the most useful tools.} of the variable $k$ along the imaginary axis between the origin and the point
$im/2$, so giving rise to the so-called \it kinematical cut \rm (because independent from the potential).
Therefore, under condition (97), the function $f(\lambda,k,r)$ is analytically defined in the region $\Im
k<m/2$, hence $f(\lambda,-k,r)$ in the region $\Im k>-m/2$. Furthermore, for Yukawian potentials of the
type\begin{equation}U(r)=\int_m^{\infty}e^{-\mu r}c(\mu)d\mu,\end{equation}it is possible to prove that Jost
solution $f(\lambda,k,r)$ is, for every $\lambda$ arbitrarily fixed, everywhere analytic in the variable $k$,
except in the branch point in $k=0$ and in a cut in the positive imaginary semi-axis made between $im/2$ and
$i\infty$, and said to be the \it dynamical cut\footnote{For certain types of potential, like the Yukawian one
(see later), such a dynamical cut may degenerate into a series of poles which are usually said to be \it false
poles \rm of the $S$ matrix, and that, in general, are poles of $f_l(k)$ in the region $\Im k>0$ where, usually,
$f_l(k)$ is not analytic. Nevertheless, we are interested in the analytical properties of the scattering
amplitude $f(k,\theta)$ that, in potential theory, does not have such a dynamical cut in the region $\Im k>0$,
so that every pole of $f(k,\theta)$ is there always associated with a bound state having a definite value of $l$
(see later).}. \rm Therefore, to sum up, the Jost solution is defined by the boundary condition (88), it is a
holomorphic function in the finite plane of $\lambda$ (at least, in the strip $|\Re\lambda|<\epsilon/2$), is an
even function in $\lambda$ and an entire function in $\lambda^2$. As regard its dependence on $k$, such a
function is analytic in the region $\Im k<0$ and continuous in the real axis except a polydromy point in the
origin which becomes a pole of order $l$ for physical values of $\lambda=l+1/2$. It is analytic in the region
$\Im k<m/2$ if the potential verifies (98); in particular, if $m=0$, it turns out to be continuous in the real
axis (except $k=0$), while if $m=\infty$, the unique finite singularity is the point $k=0$. For Yukawian
potentials, it is analytic in the whole of the plane $k$ cut between the origin and $i\infty$; for physical
angular momentum values, such a cut reduces to the dynamical one between $im/2$ and $i\infty$, plus a pole in
$k=0$. Moreover, thanks to (89), $f(\lambda,k,r)$ is, for real $\lambda$ and pure imaginary $k=i\tau$, to a real
function of $\lambda$ and $\tau$, so that, due to the Schwarz reflection principle, it has the following
Hermiticity property
\begin{equation}f^*(\lambda,k,r)=f(\lambda^*,-k^*,r)\end{equation}that, for physical values of the angular
momentum, reads as\begin{equation}f_l^*(k,r)=f_l(-k^*,r).\end{equation}Therefore, in the above mentioned
regions, the four fundamental solutions to (71), that is to say, the two regular solutions
$\phi(\pm\lambda,k,r)$ (which are also functionally independent for $\lambda\neq 0$) and the two Jost solutions
$f(\lambda,\pm k,r)$ (which are also functionally independent for $k\neq 0$), are all well-defined, so that the
ones are expressible as linear combination of the remaining others. We shall consider the regular
$\phi(\lambda,k,r)$ as a linear combination of the Jost solutions $f(\lambda,k,r)$ because, through it, we shall
be able to have a link with the $S$ matrix elements. To be precise, we consider the following combination
\begin{equation}\phi(\lambda,k,r)=\frac{1}{2ik}[f_{J}(\lambda,k)f(\lambda,-k,r)-f_{J}(\lambda,-k)
f(\lambda,k,r)]\end{equation}where\begin{equation}\begin{split}f_l(k)\equiv
f_J(\lambda,k)=&w[f(\lambda,k,r),\phi(\lambda,k,r)]=\\=&f(\lambda,k,r)\phi'(\lambda,k,r)-f'(\lambda,k,r)\phi(\lambda,k,r)
\end{split}\end{equation}is the Wronskian between $f(\lambda,k,r)$ and $\phi(\lambda,k,r)$, which define a new function
said to be the \it Jost function. \rm From (101), it is possible to prove that the asymptotic behavior of the
regular solution is given by
\begin{equation}\phi(\lambda,k,r)\underset{r\rightarrow\infty}{\sim}
\frac{1}{2ik}[f_{J}(\lambda,k)e^{i\vec{k}\cdot\vec{r}}-f_{J}(\lambda,-k)e^{-i\vec{k}\cdot\vec{r}}].\end{equation}
Moreover, by means of the boundary behaviors of the regular solutions before considered, it will be possible to
attain the following expressions for such a Jost function\footnote{Often, the expression $f(0,k)$ is also
considered as a Jost function (see (Umezawa \& Vitiello 1985, Chapter 7, Section 7.8)).}, namely
\begin{equation}\begin{split}f(\lambda,k)=&2\lambda\lim_{r\rightarrow 0}r^{\lambda-1/2}
f(\lambda,k,r)\qquad
(\Re\lambda>0),\\\\f(\lambda,k)=&2ik\lim_{r\rightarrow\infty}e^{-i\vec{k}\cdot\vec{r}}\phi(\lambda,k,r)\qquad
(\Im k<0),\end{split}\end{equation}thanks to which it will be possible to infer the properties of $f(\lambda,k)$
from the boundary behaviors and the properties of the regular solutions. For instance, the analyticity
properties of the Jost function, in the variable $\lambda$, are the same of those of the regular solution
$\phi(\lambda,k,r)$, whereas the analyticity properties of the Jost function, in the variable $k$, are the same
of those of the Jost solutions $f(\lambda,k,r)$. In particular, we have the following Hermiticity condition
$f^*(\lambda,k)=f(\lambda^*,-k^*)$ (which holds for real values of $U(r)$ and $r$) that, for physical values of
the angular momentum, implies $f^*_l(k)=f_l(-k^*)$. In absence of interaction, from the free expressions for
regular and Jost solutions, we deduce the following free Jost function
\begin{equation}\begin{split}f_0(\lambda,k)=&\sqrt{2/\pi}2^{\lambda}\Gamma(1+\lambda)e^{-i(\lambda-1/2)\frac{\pi}{2}}
k^{-\lambda+1/2}=\\\\=&2\lambda\lim_{r\rightarrow 0}r^{\lambda-1/2}f_0(\lambda,k,r).\end{split}\end{equation}

We have above considered the regular solution $\phi(\lambda,l,k)$ that, for physical values of $\lambda=l+1/2$,
that is to say, for semi-integer values of $\lambda$, it vanishes in the origin of the reference frame as
$r^{l+1}$, so that it turns out to be proportional to the physical solution $\chi_l(k,r)$, that is to say
$\phi(\lambda,k,r)\equiv\phi_l(k,r)=C_l(k)\chi_l(k,r)$, where $\chi_l$ is the so-called physical solution, in
the origin, to (67), defined by (69) that here we re-write in the following form
\begin{equation}\chi_l(k,r)\underset{r\rightarrow\infty}{\sim}\frac{e^{i\delta_l}}{2ik}[e^{i(\delta_l-l\pi/2)}
e^{i\vec{k}\cdot\vec{r}}-e^{-i(\delta_l-l\pi/2)}e^{-i\vec{k}\cdot\vec{r}}]\end{equation}so that we have the
following asymptotic behavior
\begin{equation}\phi_l(k,r)\underset{r\rightarrow\infty}{\sim}\frac{\tau_l(k)}{2ik}[e^{i(\delta_l-l\pi/2)}
e^{i\vec{k}\cdot\vec{r}}-e^{-i(\delta_l-l\pi/2)}e^{-i\vec{k}\cdot\vec{r}}],\end{equation}having put
$\tau_l(k)\equiv C_l(k)e^{i\delta_l}$. Considering complex values of $l=\lambda-1/2$ and of $k$, we re-write
(107) as follows
\begin{equation}\begin{split}\phi_l(k,r)\underset{r\rightarrow\infty}{\sim}&\frac{\tau(\lambda,k)}{2ik}
[e^{i(\delta(\lambda,k)-(\lambda-1/2)\pi/2)}e^{i\vec{k}\cdot\vec{r}}-\\&+e^{-i(\delta(\lambda,k)-(\lambda-1/2)\pi/2)}
e^{-i\vec{k}\cdot\vec{r}}],\end{split}\end{equation}and comparing this asymptotic behavior with the one given by
(103), we have the following identities
\begin{equation}\begin{split}&f(\lambda,k)=\tau(\lambda,k)e^{i(\delta(\lambda,k)-(\lambda-1/2)\pi/2))},\\\\
&f(\lambda,-k)=\tau(\lambda,k)e^{-i(\delta(\lambda,k)-(\lambda-1/2)\pi/2))},\end{split}\end{equation}which can
be considered as a parametrization of the Jost functions $f(\lambda,\pm k)$. Making the ratio between the two
formulas of (109), we see that the function $S(\lambda,k)\equiv e^{2i\delta(\lambda,k)}$ is linked to the Jost
functions $f(\lambda,\pm k)$ by the simple following relation
\begin{equation}S(\lambda,k)=e^{i(\lambda-1/2)\pi}\frac{f(\lambda,k)}{f(\lambda,-k)}\end{equation}that, for
physical values of $l$, reduces to
\begin{equation}S_l(k)=(-1)^{l}\frac{f_l(k)}{f_l(-k)}\end{equation}which represents the analytical continuation,
to complex values of $k$ and $l$, of the elements $S_l(k)$ of the $S$ matrix. Therefore, the $S$ matrix elements
$S(\lambda,k)$, as functions of the complex variables $\lambda$ and $k$, are determined by the Jost functions
$f(\lambda,\pm k)$. In particular, we are concerned with the properties of $S(\lambda,k)=S(l+1/2,k)\equiv
S_l(k)$ for fixed physical values of $l$ (and arbitrary $k$), and for fixed arbitrary values of $k$ (and
arbitrary $\lambda$), as well as in their implications for scattering amplitudes. As concern some first
analyticity properties of $S(\lambda,k)$, we may immediately say that $S(\lambda,k)$ is, for each fixed
$\lambda$ with $\Re\lambda>0$, a meromorphic function of $k$ at least in the region $|\Im k|<m/2$ (in the
hypothesis that $U(r)$ goes to zero as $r\rightarrow\infty$ at least like $e^{-mr}$) except the point $k=0$,
while $S(\lambda,k)$ is, for each fixed $k$ (in particular, for physical values of $k$), a meromorphic function
of $\lambda$ at least in the region $\Re\lambda>0$. Considering $S_l(k)$ as a function of the complex variable
$k$ with a fixed $l$ equal to one of its possible physical value, we have that (111) holds, from which it
follows that, in the $k$ plane, there may be two types of singularities: the first type corresponding to zeros
of $f_l(-k)$, the second type corresponding to the poles of $f_l(k)$ in the region $\Im k>0$ (or $\Im k>m/2$).
But what we want to highlight in this regard is the main fact that the zeros of the Jost function, hence some
poles of $S_l(k)$, are determined by the nature of the dynamical system involved in the collision process, hence
by its potential. For bound states, having a certain angular momentum $l$ and defined by a radial wave function
regular in the origin of the reference frame, say $\phi_l(k,r)$, we have that the latter is a square-integrable
function (condition characterizing a bound state that, physically, corresponds to a particle which is relegated
near the scattering center) only for $k$ not real but purely imaginary, that is to say, for $k=i\kappa$ with
$\kappa\in\mathbb{R}^+$ (take into account the symmetry of $\phi_l(k,r)$ with respect to $k\rightarrow -k$).
From (103), it follows that
\begin{equation}\phi_l(i\kappa,r)\underset{r\rightarrow\infty}{\sim}-\frac{1}{2k}[f_l(i\kappa)e^{-\kappa r}-
f_l(-i\kappa)e^{\kappa r}]\end{equation}so in order to have a bound state for $k=i\kappa$ with energy
$E=-\hbar^2\kappa^2/2\mu$, it is necessary that the coefficient of $e^{\kappa r}$ is zero, that is
\begin{equation}f_l(-i\kappa)=0,\end{equation}so that
\begin{equation}\phi_l(i\kappa,r)\underset{r\rightarrow\infty}{\sim}-\frac{1}{2k}f_l(i\kappa)e^{-\kappa r}\end{equation}
which is normalizable, whence it describes a bound state. Thus, (113) provides a necessary and sufficient
condition for the existence of a bound state having angular momentum $l$ and energy $E=-\hbar^2\kappa^2/2\mu$,
that is, a bound state corresponds to a zero of the Jost function in the half-plane $\Im k<0$ on the negative
imaginary semi-axis $\Re k=0$. In such a case, the wave function $\phi_l(i\kappa,r)$ describing such a state is
a real function because, taking into account the parity symmetry $k\rightarrow -k$, we have
\begin{equation}\phi_l^*(i\kappa,r)=\phi_l(-i\kappa,r)=\phi_l(i\kappa,r).\end{equation}Therefore, as regard the
$S$ matrix elements, the bound states manifest themselves both as poles of $S_l(k)$ in $\Im k>0$, due to the
zeros of $f_l(-k)$, and as zeros (placed symmetrically with respect to the poles) of $f_l(k)$ in $\Im k<0$ and
placed into the semi-axis $\Re k=0$. It is also possible to prove that, for bound states, the latter are simple,
so that the corresponding poles of $S_l(k)$ are simple too. If one denotes with $\dot{f}_l(k)$ the first
derivative of $f_l(k)$ with respect to $k$, then it is possible to prove that, for bound states, the following
relation holds
\begin{equation}\int_0^{\infty}[\phi_l(k_0,r)]^2dr=\frac{1}{4ik_0^2}\dot{f}_l(k_0)f_l(-k_0),\end{equation}while,
an expression for the residues of $S_l(k)$, with respect to the variable $k=i\kappa\ (\kappa>0)$, into the poles
corresponding to bound states, is as follows
\begin{equation}\displaystyle R_l(i\kappa)=-i(-1)^l\frac{N_l^2(\kappa)}
{\displaystyle\int_0^{\infty}\phi_l^2(i\kappa,r)dr}\end{equation}where $N_l(k)\equiv -f_l(i\kappa)/2\kappa$ is
the normalization constant of the asymptotic wave function given by (114). It follows that such a residue is
purely imaginary, while, if one uses the variable $E=\hbar^2k^2/2\mu$ instead of $k$, in regard to a bound state
corresponding to $k=i\kappa$, the residue itself computed with respect to $E=E_{\kappa}=-\hbar^2\kappa^2/2\mu$
is given by\begin{equation}r(E_{\kappa})=\lim_{E\rightarrow
E_{\kappa}}(E-E_{\kappa})S_l(k)=\frac{i\kappa\hbar^2}{\mu}R_l(i\kappa)\end{equation}from which it turns out to
be real.

Besides\marginpar{\tiny\it Bound states, resonances, virtual states} the above mentioned poles in the half-plane
$\Im k>0$, with respect to the parity symmetry $k\rightarrow -k$, $S_l(k)$ has further poles also in the
half-plane $\Im k<0$, corresponding to the possible zeros of $f_l(-k)$ in such a half-plane, where, in general,
it is not analytic. In such poles, the wave functions $\phi_l(k,r)$ have an explosive behavior and, therefore,
do not represent bound states. In the region $\Im k<0$, we may have poles either with $\Re k\neq 0$ and $\Re
k=0$. In the first case, if $k=k_0$ is a pole of $S_l(k)$ in the region $\Im k<0$, the case $\Re k_0\neq 0$
(which cannot to take place for bound states due to the related square-summable finiteness condition) implies
the existence of pairs of poles symmetrically placed with respect to the axis $\Re k=0$. The poles of $S_l(k)$
as zeros of $f_l(-k)=0$ in the region $\Im k<0$ with $\Re k\neq 0$, are usually called \it resonance, \rm while
the poles of $S_l(k)$ as zeros of $f_l(-k)=0$ in the same region but with $\Re k=0$, are said to be \it virtual
states \rm (or \it antibound states \rm or \it redundant zeros\footnote{The redundant zeros were first discussed
by S.T. Ma in (Ma 1946; 1947a,b).}\rm). For a resonance, that is to say, for a pole of $S_l(k)$ of the type
$k=h+ib$ with $b\leq 0$, we never have $b=\Im k=0$ since, if it were $f_l(-k)=0$ for $k\in\mathbb{R}$, due to
the Hermiticity condition $f_l^*(k)=f_l(-k^*)$, we would have too $f_l^*(-k)=f_l(k)=0$, so that the wave
function $\phi_l(k,r)$ would turn out to be zero as well. We give the name resonance to such a type of pole
because, when $b$ is small, its presence is manifested in the scattering amplitude through the appearance, in
the $l$-th wave function component, of a resonance, centered around an energy value $E=E_0=\hbar^2h^2/2\mu$ and
having a width $\Gamma$ proportional to $|b|$. This last reasoning holds too for all the poles of $S_l(k)$ in
$\Im k<0$ with $\Re k\neq 0$, for every physical value of $l$. Instead, an antibound state meant as a zero of
$f_l(-k)$ in the region $\Im k<0$ with $\Re k=0$, say $k=ib$ with $b<0$, may manifest with an exaltation of the
scattering amplitude at the threshold, if $|b|$ is very small, even if this last phenomenon is also due to the
presence of a bound state. For what follows, in studying analytical properties of $S_l(k)$ for physical values
of $l$, it is better to choose $E\equiv \hbar^2k^2/2\mu$ as a fundamental variable rather than
\begin{equation}k=\sqrt{E}\sqrt{2\mu/\hbar^2}.\end{equation}Therefore, to get full variability of $k$, $E$ must
variate over all the two sheeted Riemann surface of $\sqrt{E}$, made by two sheets with branch point at the
origin except for a cut on both sheets extending along $\Re k\leq 0$. In such a case, it is a common practice to
cut the plane of the complex variable $E$ along the positive real semi-axis, calling \it physical sheet \rm(or
\it first sheet\rm) of such a Riemann surface the one corresponding to the positive determination of $\sqrt{E}$
(hence that corresponding to $\Im k>0$), and \it non-physical sheet \rm(or \it second sheet\rm) the one
corresponding to the negative determination of $\sqrt{E}$ (hence corresponding to $\Im k<0$). Thus, for
$S_l=S_l(E)$, we have that bound states are poles of $S_l(E)$ placed on the negative real semi-axis of the
physical sheet (being, for bound states, $\Im k>0$ and $\Re k\leq 0$); the resonances are poles of $S_l(E)$
which lie on the non-physical sheet (being $\Im k<0$) and the nearer they are to the cut, the stronger their
effects will be on the scattering cross-section; and the antibound states are poles of $S_l(E)$ which lie on the
negative real semi-axis of the non-physical sheet (being $\Im k<0$ and $\Re k\leq 0$) and the nearer they are to
the origin of the reference frame, the stronger their effect will be on the scattering cross-section in
threshold; in this latter case, moreover, we have trajectories which behave for a while as bound states but
finally break-up into two particles moving apart (see (Mackey 1978, Section 21)). Therefore, it is meaningful to
speak of either \it close \rm polar singularities with respect to the physical region $E>0$, represented by the
cut, and \it far \rm polar singularities, the first ones being those that mainly contribute to determine the
form of the scattering cross-section. Furthermore, if the pole of $S_l(k)$ is at the point $k=a+ib$, then it
represents our unstable bound state of energy $a$ and angular momentum (or spin) $l$; the larger $b$ is, the
shorter the particle lifetime will be, and when the lifetime is so short that the bound state decomposes before
it can travel a measurable distance, then we have a resonance. Of course, there is no sharp distinction between
resonances and unstable particles. It is interesting that resonances and unstable particles not only have
energies and angular momenta but behave like stable particles in other respects. One can even discuss their
interactions with one another (see (Mackey 1978, Section 21)).

Now, we\marginpar{\tiny\it Dispersion relations} consider the scattering amplitude $f(k,\theta)$ (as, for
instance, given by (63)) as a function of the energy $E=\hbar^2k^2/2\mu$ and of the scattering angle $\theta$,
hence denoted by $f(E,\theta)$. We may consider $E$ as a complex variable. For simplicity's sake, we consider
natural units, so that $\hbar=2\mu=1$, whence $k=\sqrt{E}$. Therefore, considering $f(E,\theta)$ as a function
of $E$ at each fixed value of $\theta$, from (63) we have the following partial wave
development\begin{equation}f(E,\theta)=\sum_0^{\infty}(2l+1)a_l(k)P_l(\cos\theta)\end{equation}
where\begin{equation}a_l(k)=\frac{1}{2ik}(S_l(k)-1)\end{equation}and $k=\sqrt{E}$. Therefore, the singularities
of the partial amplitudes $a_l$ are the same of $S_l(k)$, and since it is possible to prove that the dynamical
cut of $S_l(k)$ does not appear in the scattering amplitude, we have that the unique singularities of
$f(E,\theta)$, in the variable $E$, are poles corresponding to the possible bound states, resonances and virtual
(or antibound) states, besides the possible polydromy points of $\sqrt{E}$, i.e., origin and point at infinity.
We restrict to consider $f(E,\theta)$ over the physical sheet of $\sqrt{E}$, where the unique possible
singularities, for what has been said above, are the points of the cut along the positive real semi-axis as well
as possible bound states having a definite angular momentum $l$ and placed into the non-positive real semi-axis.
Moreover, the singularities of the scattering amplitude $f(E,\theta)$ on the physical sheet coincide with the
spectrum of the Hamiltonian of the dynamical system constituted by the interacting particles, including the
allowed bound states as poles of $f(E,\theta)$ for $\Re E<0$, which are in a finite number for every potential
decreasing faster than $1/r^2$ (hence belonging to the discrete spectrum), and all the states of the cut of
$f(E,\theta)$ which correspond to all values of $\Re E>0$ (hence belonging to the continuous spectrum). Then, as
$E\rightarrow\infty$ by means of any direction along the physical sheet, it is possible to prove that
$f(E,\theta)\underset{E\rightarrow\infty}{\rightarrow}f_B(\theta)$ in a uniform manner, being $f_B(\theta)$ the
Born scattering amplitude obtained considering the Born approximation $(22)_2$ which holds when high energy
values are called into question. To this point, it is important to show that the scattering amplitude
$f(E,\theta)$ undergoes a certain energy dispersion relation computed with respect to a fixed value of the angle
$\theta$. To get it, we consider the Cauchy integral representation of $f(E,\theta)-f_B(\theta)$ extended along
a closed path of the $E'$ plane this last being cut along the $\Re E'>0$, having poles along $\Re E'\leq 0$,
surrounding the point $E$ but no one of the singularities of the integrating function. Thanks to Cauchy theorem,
this integration path, say $\Upsilon$, may be homotopically deformed into a bigger circle, with radius
$R\rightarrow\infty$, closely embracing the cut $\Re E'>0$ plus small neighborhoods each of which centered
around every pole of $\Re E'\leq 0$. As we have said before, $f(E,\theta)-f_B(\theta)$ uniformly tends to zero,
so that the contribution of the integral\begin{equation}f(E,\theta)-f_B(\theta)=\frac{1}{2\pi
i}\oint_{\Upsilon}\frac{f(E',\theta)-f_B(\theta)}{E'-E}dE',\end{equation}along this circle, goes to zero as
well, so that we may write\begin{equation}\begin{split}f(E,\theta)=&f_B(\theta)+\\+&\frac{1}{2\pi
i}\int_{0}^{\infty}\frac{f(E'+i\epsilon,\theta)-f(E'-i\epsilon,\theta)}{E'-E}dE'+
\sum_n\frac{y_n(\theta)}{E-E_n},\end{split}\end{equation}the first contribution being extended to that part of
the path $\Upsilon$ embracing $\Re E'>0$ along a narrow strip of width $\epsilon\rightarrow 0$, whilst the
second one comprehends each contribution given by the neighborhood of every pole of $\Re E'\leq 0$, where
$y_n(\theta)=\mbox{\rm Res}\{f(E,\theta)\}_{E=E_n}$, for each $n$, is the residue of the scattering amplitude in
the pole $E=E_n$ (which represents a bound state having a definite angular momentum $l$) and the summation is
over all the allowed bound states. For $\theta\in\mathbb{R}$, we have $f^*(E,\theta)=f(E^*,\theta)$, whence
$f(E-i\epsilon,\theta)=f^*(E+i\epsilon,\theta)$, so that (123) reduces to
\begin{equation}f(E,\theta)=f_B(\theta)+\frac{1}{\pi}\int_{0}^{\infty}
\frac{\Im f(E',\theta)}{E'-E}dE'+ \sum_n\frac{y_n(\theta)}{E-E_n}\end{equation}where $\Im f(E',\theta)$ is
evaluated upon the upper boundary of the cut. The dispersion relation (124) is due the classical form of the \it
Khuri's dispersion relation \rm (see (Khuri 1957)). This last expression is nothing but a dispersion relation
for the scattering amplitude which provides all the possible singularities and characteristics: namely, the cut
along the positive real semi-axis and the discontinuities of $f(E,\theta)$ in it (which are $2i\Im
f(E,\theta)$), as well as the the poles corresponding to bound states for $E=E_n\ (\Re E_n<0)$ and the related
residues $y_n(\theta)$; furthermore, it provides the behavior of $f(E,\theta)$ as $E\rightarrow\infty$, being
$f(E,\theta)\rightarrow f_B(\theta)$. As regard, then, the residues $y_n(\theta)$, these can be evaluated
through the corresponding functions $S_l(k)$ as follows
\begin{equation}y_n(\theta)=(2l+1)\frac{r_l(E_n)}{2\sqrt{-E_n}}P_l(\cos\theta)\in\mathbb{R}\end{equation}
where $r_l(E_n)$ is the residue of $S_l(k)$ in $E=E_n$, basically given by (117) and (118). Nevertheless, the
(124) not always turns out to be useful because it might be quite difficult to have information on $\Im
f(E,\theta)$, so that, through the optical theorem (57) in the energy $E$ with respect to $\theta=0$, another
formula is often used instead of (124), namely the following one
\begin{equation}\Re f(E,0)=f_B(0)+\frac{P}{4\pi^2}\int_{0}^{\infty}
\frac{\sqrt{E'}\sigma_t(E')}{E'-E}dE'+ \sum_n\frac{y_n(0)}{E-E_n}\end{equation}which is said to be the \it
forward dispersion relation \rm of the scattering amplitude\footnote{We have
$1/(E'-E-i\epsilon)=P/(E'-E)+i\pi\delta(E'-E)$.}, connecting the real part of $f(E,0)$ with an integral over its
imaginary part as well as with the position and residues of the poles. It was provided by D. Wong and N.N. Khuri
in 1957.

We reconsider\marginpar{\tiny\it Regge poles and trajectories} $S(\lambda, k)$, as given by (110), as a function
of the complex variable $\lambda$, for any $k\in\mathbb{R}^+$ arbitrarily fixed. For $k\in\mathbb{R}^+$
(positive energy), we have the following Hermiticity condition $S^*(\lambda,k)=[S(\lambda^*,k)]^{-1}$ which
implies, as regard the scattering phases $\delta$ of $S(\lambda,k)=e^{2i\delta(\lambda,k)}$, the relation
$\delta^*(\lambda,k)=\delta(\lambda^*,k)$, so that they are real when $\lambda$ is also real. From what has been
said above, $S(\lambda,k)$ is a meromorphic function in the variable $\lambda$ at least in the region
$\Re\lambda>0$, where $f(\lambda,\pm k)$ are both analytic when $k$ is real, so that, in the region
$\Re\lambda>0$, due to zeros of $f(\lambda,-k)$, poles of $S(\lambda,k)$ may exist, which are universally known
as \it Regge poles\rm; they are confined in the region $\Im\lambda>0$ and there exist independently of the
potential $U(r)$, that is, their existence has a pure kinematical nature. If one considers further restriction
on the potential $U(r)$, then the localization of the Regge poles may be more accurately identified. For
instance, for purely imaginary values of $r$, if $|U(i\rho)|<M/\rho^2$ with $M\in\mathbb{R}^+$, then the Regge
poles are confined, in the half-plane $\Re\lambda>0$ and $\Im\lambda>0$, within the region delimited by the
hyperbole branch $\Re\lambda\cdot\Im\lambda<M/2$ and by their asymptotes $\Re\lambda=0$ and $\Im\lambda=0$.
Instead, if $|U(i\rho)|<N/\rho$ with $N\in\mathbb{R}^+$, then the Regge poles are confined into the region
$\Re\lambda<N/k$, whereas, for fixed values of $k\in\mathbb{R}^+$, the half-plane on the right hand side of the
line $\Re\lambda=N/k$, is a region in which $S(\lambda,k)$ is holomorphic. Moreover, for Yukawian potentials
verifying $|U(i\rho)|<M/\rho^2$, it is possible to prove that the real part of the Regge poles is upperly
limited by $\Re\lambda=M'$ so that the half-plane $\Re\lambda>M'$ is a holomorphic zone for $S(\lambda,k)$. We
have also need to know the behavior of $S(\lambda,k)$ as $|\lambda|\rightarrow\infty$ in the region
$\Re\lambda>0$. To this end, it is possible to prove that the following relation holds
\begin{equation}\lim_{|\lambda|\rightarrow\infty}S(\lambda,k)=1\end{equation}at least into certain regions of
the $\lambda$ plane. Such a relation may be also deduced by some estimates like the following ones. If
$S(\lambda,k)=e^{2i\delta(\lambda,k)}$, then
\begin{equation}|S(\lambda,k)-1|=O(\lambda^{-1/2}e^{-\alpha\lambda})\end{equation}for potentials $|U(r)|<ce^{-mr}/r$ such
that $\cosh\alpha=1+m^2/2k^2$. For complex values of $\lambda$, at least when the real part of the poles of
$S(\lambda,k)$ is upperly bounded, for $|\lambda|$ great enough and such that $|\mbox{\rm arg}\ \lambda|<\pi/2$,
we have\begin{equation}|S(\lambda,k)-1|<ce^{-\alpha\Re\lambda},\end{equation}so that $S(\lambda,k)-1$
exponentially tends to zero as $\lambda\rightarrow\infty$ in the region $\Re\lambda>0$. It is desirable that
many results of potential theory, when it is possible, may be extended to the relativistic case. In doing so,
with respect to natural units, it is better to replace the fundamental kinematical variables, the relative wave
vector $k$ and the scattering angle $\theta$, with the following two other variables, namely $s=E=k^2$ and the
new parameter
\begin{equation}t\equiv q^2=2k^2(1-\cos\theta),\end{equation}which respectively represent the relative energy of
the incident particle beam and the square of the transfer momentum during the scattering process. So, the
scattering amplitude $f(s,t)$, expressed in these two new variables, has a physical meaning only when
$s\in\mathbb{R}^+$ and $t\in\mathbb{R}^-\cap\ ]-4s,0[$. But, from these values, it is possible to analytically
prolong such a function, as well as its asymptotic behavior, to all complex values of $s$ and $t$, whose
analytical properties are often obtainable by means of certain dispersion relations involving such a function,
like (124), which is a dispersion relation at a fixed $\theta$ (hence at a fixed $t$), or the so-called \it
double representation \rm (in $s$ and $t$) of S. Mandelstam (see (Chew 1962)), which allows to determine the
scattering amplitude without make use of any wave equation. The asymptotic behavior of $f(s,t)$ as
$t\rightarrow\infty$ has no physical meaning in the non-relativistic context but only in the relativistic one
where the particle/antiparticle symmetry, which entails invariance with respect to the exchange between the two
variable $s$ and $t$, provides physical meaning to such a behavior. Indeed, at relativistic level, the so-called
\it substitution law \rm holds, according to which, together a diffusion process of the
type\begin{equation}a+b\rightarrow a'+b',\end{equation}it is needed to consider the following \it
cross-diffusion \rm process\begin{equation}a+\bar{a}'\rightarrow\bar{b}+b'\end{equation}where $\bar{a}$ denotes
the antiparticle of $a$, and so forth. In a few words, the substitution law states that, in a diffusion process,
an incoming particle (like $\bar{a}'$) is equivalent to its outgoing antiparticle (like $a$). The quantity
$t\equiv q^2$, is the transfer momentum into the direct channel (131) as well as the energy of the cross channel
(132). $t$ is a limited quantity for the former, being $0\leq t\leq 4k^2$, whilst isn't for the cross channel
for which $t$ may take any value in $\mathbb{R}^+$, so that it is possible to consider the asymptotic behavior
of the scattering amplitude
\begin{equation}f(k,\theta)=f\Big[\sqrt{E},\arccos\big(1-\frac{t}{2k^2}\big)\Big]\equiv F(E,t)\end{equation} as
$t\rightarrow\infty$. If we work at a fixed value of energy, that is to say $k\in\mathbb{R}^+$ is arbitrarily
fixed, then, by (130), it follows that $t\rightarrow\infty$ is equivalent to $\cos\theta\rightarrow -\infty$, so
that it is needed to search for a representation of the scattering amplitude $f(k,\theta)$, which is valid for
physical values of $\theta$ and that may be analytically prolonged in a open region of the plane of $\cos\theta$
in which it is possible bringing $\cos\theta$ to $-\infty$.

In pursuing this, we start from the following partial wave expansion of the scattering amplitude
\begin{equation}f(k,\theta)=\sum_{l=0}^{\infty}(2l+1)a_l(k)P_l(\cos\theta)\end{equation}which defines
$f(k,\theta)$ as an analytic function of $\cos\theta$ in that region of the plane of the complex variable
$z\equiv\cos\theta$ in which the series at the right hand side of (134), converges. Such a convergence region
depends on the $l$-th term of the series as $l\rightarrow\infty$, so that we should study the behavior of
$a_l(k)$ and $P_l(\cos\theta)$ as $l\rightarrow\infty$. To this end, we have the following asymptotic relations
\begin{equation}P_l(\cos\theta)=O\Big(\frac{e^{l|\Im\theta|}}{\sqrt{l}}\Big),\qquad |a_l(k)|=
O\Big(\frac{e^{-\alpha l}}{\sqrt{l}}\Big)\end{equation}where $\alpha$ is a positive real constant as above
defined by $\cosh\alpha=1+m^2/2k^2$, so that the $l$-th term of the series (134) exponentially goes to zero as
$l\rightarrow\infty$ in the region $|\Im\theta|<\alpha$ which, therefore, identifies the convergence region of
the series (134) in the $\theta$ plane. To determine the strip $|\Im\theta|<\alpha$ in the plane of
$z\equiv\cos\theta$, let us set $x=\Re z$, $y=\Im z$, $\theta_1=\Re\theta$ and $\theta_2=\Im\theta$, whence
\begin{equation}z\equiv x+iy=\cos\theta\equiv\cos (\theta_1+i\theta_2)=\cos\theta_1\cosh\theta_2
-i\sin\theta_1\sinh\theta_2\end{equation}from which it follows that
\begin{equation}x=\cos\theta_1\cosh\theta_2,\qquad y=-\sin\theta_1\sinh\theta_2,\end{equation}so the
boundary of the convergence region of (134) is the boundary of the strip $|\Im\theta|<\alpha$, that is to say
$\theta_2\equiv\Im\theta=\pm\alpha$, and inserting these values of $\theta_2$ into (137), hence eliminating
$\theta_1$, we get the equation of the boundary of the convergence region of (134) in the plane of
$z\equiv\cos\theta$, that is the ellipse
\begin{equation}\frac{x^2}{\cosh^2\alpha}+\frac{y^2}{\sinh^2\alpha}=1,\end{equation}said to be the \it Lehmann
ellipse, \rm (see (Lehmann 1958)) having major radius $\cosh\alpha=1+m^2/2k^2$, minor radius
$\sinh\alpha=\sqrt{\cosh^2\alpha-1}$ and foci in the points $\cos\theta=\pm 1$, the extremes of the physical
region related to (134), which identifies the largest ellipse in the $\cos\theta$ plane where the scattering
amplitude is analytic. It is possible to prove that, through (137), the strip $|\Im\theta|<\alpha$ is the
interior of the region identified by (138), so that the convergence region of (134), in the plane of
$\cos\theta$, is the interior of the Lehmann ellipse. In the plane $t=2k^2(1-\cos\theta)$, the Lehmann ellipse
remains an ellipse, with foci in $t=0$ and $t=4k^2$, the point $t=-m^2$ is the extreme of the major radius
corresponding to $\cos\theta=1+m^2/2k^2$, its two semi-axes depends on the energy through the above expressions
of $\cosh\alpha$ and $\sinh\alpha$, and the more the energy grows up, the smaller and the more restricted such
an ellipse becomes, at last reducing to the physical region given by the segment $[-1,1]\subset\Re z$ of the
plane of $z=\cos\theta$ as $k\rightarrow\infty$. Therefore, (134) is a representation of the scattering
amplitude which converges within the Lehmann ellipse where it turns out to be an analytic function of
$\cos\theta$. Nevertheless, this representation is unable to study the behavior of the scattering amplitude as
$t\rightarrow\infty$ or equivalently as $\cos\theta\rightarrow -\infty$ that, however, may be suitably
re-written to accomplish this purpose transforming the series into an integral over the complex plane of the
above considered variable $\lambda=l+1/2$, in such a manner to have an analytic continuation of $f(k,\theta)$
over an open domain of the plane of $\cos\theta$; such a particular integral transformation is usually called
\it Watson-Sommerfeld transformation. \rm To introduce this, we need to consider the Legendre function
$P_l(z)\equiv P_{\lambda-1/2}(z)$ for arbitrary complex values of $\lambda$, which is an even entire function of
$\lambda^2$. For $z=\cos\theta$, we have the following estimate
\begin{equation}|P_{\lambda-1/2}(\cos\theta)|\leq c|\sin\theta|^{-1/2}|\lambda|^{-1/2}
e^{|\Re\theta||\Im\theta|-\Im\theta\cdot\Re\lambda}.\end{equation}Now, we consider the function
\begin{equation}G(\lambda)\doteq\lambda\frac{S(\lambda,k)-1}{\cos\pi\lambda}P_{\lambda-1/2}(-\cos\theta)\end{equation}
which is a meromorphic function of $\lambda$ in $\Re\lambda>0$, with poles in the zeros of $\cos\pi\lambda$ (in
correspondence to the physical values of $\lambda=l+1/2$) plus further possible poles in $\Im\lambda>0$ given by
$S(\lambda,k)$. The residue of $G(\lambda)$ in a generic pole in the variable $\lambda=l+1/2$, with $l$ having
physical values, is\begin{equation}\mbox{\rm
Res}\{G(\lambda)\}_{\lambda=l+1/2}=-\frac{1}{2\pi}(2l+1)[S_l(k)-1]P_l(\cos\theta).\end{equation}If one defines
the integral\begin{equation}\int_{\mathcal{C}}G(\lambda)d\lambda\end{equation}where $\mathcal{C}$ is the path
surrounding all and only all the poles of the real semi-axis $\Re\lambda>0$, then, from the residue theorem, we
have\begin{equation}\int_{\mathcal{C}}G(\lambda)d\lambda=-2\pi i\sum_{l=0}^{\infty} \mbox{\rm
Res}\{G(\lambda)\}_{\lambda=l+1/2}=-2kf(k,\theta)\end{equation}whence we have the following integral
representation\begin{equation}f(k,\theta)=-\frac{1}{2k}\int_{\mathcal{C}}\lambda\frac{S(\lambda,k)-1}
{\cos\pi\lambda}P_{\lambda-1/2}(-\cos\theta)d\lambda.\end{equation}We would like to homotopically deform the
path $\mathcal{C}$ in such a manner it coincides with the imaginary axis. First, we suppose do not exist poles
of $S(\lambda,k)$ in $\Re\lambda>0$. We notice that, as $|\lambda|\rightarrow\infty$ along any direction of the
half-plane $\Re\lambda>0$, from (139) with $\cos\theta\rightarrow -\cos\theta=\cos(\pi-\theta)$, it follows
\begin{equation}\Big|\frac{\lambda P_{\lambda-1/2}(-\cos\theta)}{\cos\pi\lambda}\Big|\leq c(\sin\theta)^{-1/2}
|\lambda|^{1/2}e^{-|\Re\theta\cdot\Im\lambda|+\Im\theta\cdot\Re\lambda},\end{equation}and, if
$|S(\lambda,k)-1|<ce^{-\alpha\Re\lambda}$, then, from (144), it follows that
\begin{equation}|G(\lambda)|<c|\lambda|^{1/2}e^{-|\Re\theta\cdot\Im\lambda|+(\Im\theta-\alpha)\Re\lambda}\end{equation}
whence, in turn, it follows that $G(\lambda)$ goes to zero uniformly as $|\lambda|\rightarrow\infty$ along any
direction in $\Re\lambda>0$, provided that $|\Im\theta|<\alpha$, which is a condition surely satisfied within
Lehmann ellipse. Therefore, in these conditions (above all, in absence of poles of $S(\lambda,k)$), the
deformation of $\mathcal{C}$ to the imaginary axis $\Im\lambda$ is allowed, so obtaining the following integral
representation of the scattering amplitude
\begin{equation}\begin{split}f(k,\theta)=&-\frac{1}{2k}\int_{-i\infty}^{i\infty}G(\lambda)d\lambda =\\
=& -\frac{1}{2k}\int_{-i\infty}^{i\infty}\lambda\frac{S(\lambda,k)-1}{\cos\pi\lambda}P_{\lambda-1/2}
(-\cos\theta)d\lambda\end{split}\end{equation}which is said to be the \it Watson-Sommerfeld transform \rm
(briefly \it WS transform\rm) of $S(\lambda,k)$; it expresses the scattering amplitude $f(k,\theta)$ as a
continuous superposition of \it conical functions, \rm that is to say, Legendre functions of the type
$P_{it-1/2}(x)$ with $t\in\mathbb{R}$. Following (De Alfaro \& Regge 1965, Chapter 13, Section 13.2), the
formula (147) looks like an infinite level Breit-Wigner formula. In (147), the variables $\lambda$ and
$\cos\theta$ are conjugate of each other by means of the so-called \it Melher-Fock inversion
formulas\footnote{See (Mehler 1881) and (Fock 1943), as regard the original papers.} \rm through which it is
possible to consider $S(\lambda,k)$ in terms of $f(k,\theta)$; therefore, the properties of $f(k,\theta)$ with
respect to the variable $\cos\theta$, may be deduced by the properties of $S(\lambda,k)$ with respect to the
variable $\lambda$, and vice versa. Of course, further questions of convergence and related areas of convergence
of the integral of (147) occur: in any case, what is important is to notice that there exists a common
convergence region in which both representations (134) and (147) are the analytic continuation of each other; in
particular, the representation (147) comprises the asymptotic region $\cos\theta\rightarrow -\infty$, so that
the Watson-Sommerfeld transform is suitable for studying the behavior of the scattering amplitude as
$\cos\theta\rightarrow -\infty$, that is, as $t\rightarrow\infty$, whose properties in function of
$t=2k^2(1-\cos\theta)$ are obtainable from those in $\cos\theta$.

It is interesting to consider the Watson-Sommerfeld transform of $S(\lambda,K)$ in presence of Regge poles, that
is to say, in the case in which $S(\lambda,k)$ has a certain number of poles in the quadrant
$\Re\lambda>0,\Im\lambda>0$. If, for each fixed value of $k$, $S(\lambda,k)$ has a pole of the type
$\lambda=\alpha_n(k)+1/2$ with residue $\sigma_n=\sigma_n(k)$, then the residue of the function $G(\lambda)$,
given by (140), computed in this pole, will be
\begin{equation}\{\mbox{\rm Res}\ G(\lambda)\}_{\lambda=\alpha_n+1/2}=-(\alpha_n+1/2)\frac{\sigma_n(k)}
{\sin\pi\alpha_n}P_{\alpha_n}(-\cos\theta).\end{equation}Therefore, in applying the Watson-Sommerfeld transform
along the deformation of the original path $\mathcal{C}$ to the imaginary axis, we must further consider a
closed path which surrounds every single pole of such a type when we apply (143) in pursing this, each of which
provides the following contribution
\begin{equation}\frac{\beta_n(k)}{\sin\pi\alpha_n(k)}P_{\alpha_n(k)}(-\cos\theta)\end{equation}where
\begin{equation}\beta_n(k)=\frac{i\pi}{k}[\alpha_n(k)+\frac{1}{2}]\sigma_n(k),\end{equation}so that, in presence
of Regge poles, the Watson-Sommerfeld transform assumes the following expression\begin{equation}\begin{split}
f(k,\theta)=&-\frac{1}{2k}\int_{-i\infty}^{i\infty}\lambda\frac{S(\lambda,k)-1}{\cos\pi\lambda}P_{\lambda-1/2}
(-\cos\theta)d\lambda+\\&+\Sigma_n\frac{\beta_n(k)}{\sin\pi\alpha_n(k)}P_{\alpha_n(k)}(-\cos\theta)
\end{split}\end{equation}the summation being over all the existent Regge poles. This is the fundamental formula
to study the asymptotic behavior of the scattering amplitude $f(k,\theta)=F(E,t)$ as $\cos\theta\rightarrow
-\infty$, that is, as $t\rightarrow\infty$, which we now will carry out only in the case of a unique Regge pole
$\lambda=\alpha(k)+1/2$, in such a manner that (151) reduces to\begin{equation}\begin{split}
f(k,\theta)=&-\frac{1}{2k}\int_{-i\infty}^{i\infty}\lambda\frac{S(\lambda,k)-1}{\cos\pi\lambda}P_{\lambda-1/2}
(-\cos\theta)d\lambda+\\&+\frac{\beta(k)}{\sin\pi\alpha(k)}P_{\alpha(k)}(-\cos\theta)
\end{split}\end{equation}with $k=\sqrt{E}$ and $-\cos\theta=(t/2k^2)-1=(t/2E)-1$. Now, from the asymptotic
behavior of Legendre functions, it is possible to deduce that the integral at the right hand side of (152) goes
to zero as $t\rightarrow\infty$, which is the main reason for which it is often called the \it background
integral\rm; furthermore, it is also possible to prove that Regge pole behaves like $t^{\alpha(k)}$, so that, in
the case of a unique Regge pole with $\Re\alpha>-1/2$, this dominates the background integral, so obtaining the
following asymptotic expansion of the scattering amplitude given by (133)
\begin{equation}F(E,t)\underset{t\rightarrow\infty}{\sim}c(E,\mbox{\rm Res}\
G(\alpha(k)))t^{\alpha(k)}\qquad\mbox{(\it Regge's\ theorem\rm)}\end{equation}where $c(E,\mbox{\rm Res}\
G(\alpha(k)))$ is a constant which depends on the energy and on the residue of the Regge pole. The asymptotic
formula (153) still holds for a finite number of Regge poles in which $\alpha$ will be replaced by that pole
$\alpha_n$ with highest real part, whereas, in general, it does not hold for an infinite number of poles.
Regge's theorem is a consequence of Watson-Sommerfeld representation of scattering amplitude. Attempts to extend
toward relativistic regimes -- or, however, to high energy contexts -- the relation (153) have been made: in
this latter case, the (153), thanks as well to the substitution law, shows that the high-energy behavior of the
scattering amplitude is mainly ruled by the related Regge poles involved in the given collision process, besides
certain \it Regge cuts \rm of the quadrant $\Re\lambda>0,\Im\lambda>0$ whose existence seems to be suggested by
various physical models and that contribute as well to the asymptotic behavior of the scattering amplitude. We
finally will try to shed light on the possible physical interpretation of the Regge poles.

A single Regge pole provides the following contribution to the scattering amplitude,
\begin{equation}\frac{\beta}{\sin\pi\alpha}P_{\alpha}(-\cos\theta)\equiv f_R(k,\theta)\end{equation}where the
position of the pole $\alpha(k)+1/2$ depends on the energy $E=k^2$, while $\beta$ is a residue which, in turn,
depends on the energy as well. From (65) and (154), as well as taking into account certain properties of the
Legendre functions, it follows that the contribution of this pole to the $l$-th partial amplitude is
\begin{equation}a_l^R(k)=\frac{1}{2}\int_{-1}^1f_R(k,\theta)P_l(\cos\theta)d\cos\theta=\frac{1}{\pi}
\frac{\sin\pi\alpha}{(l-\alpha)(l+\alpha+1)}\end{equation}with $\alpha=\alpha(E),\beta=\beta(E)$, from which it
turns out that this contribution of the pole to the $l$-th partial amplitude is very relevant when
$\alpha\approx l$: for instance, if, whilst $E$ is varying, $\alpha(E)$ takes, for $E=E_0$, a value, say
$\alpha_0$, much near to an integer one, say $l$, then the $l$-th partial wave shows a resonance in
correspondence at $E=E_0$. This last argument may be also supported by the fact that (155) is closely related to
the Breit-Wigner formula (20) because it is possible to prove that it reduces to a formula of Breit-Wigner type.
In fact, we suppose that $S(\lambda,k)$ has a Regge pole for $\lambda=\alpha(E)+1/2$ and that, for $E=E_0$, let
$\alpha(E_0)=\alpha_0$ be much near to an integer value $l$, that is to say
$\alpha(E_0)=\alpha_0=l+\eta_0+i\sigma_0$ where\footnote{The Regge pole
$\alpha(E_0)=\alpha_0=(l+\eta_0)+i\sigma_0$, has positive imaginary part $\sigma_0$ because every Regge pole
lies in the region $\Im\lambda>0$.} $|\eta_0|\ll 1,\sigma_0\ll 1$. But, if $S(\lambda,k)$, for $k=k_0$ (that is
to say, for $E=E_0$), has a pole for $\lambda=\alpha(E_0)+1/2$, then the Jost function $f(\lambda,-k)$ is zero
in these values, that is $f(\alpha(E_0)+1/2,-k_0)=0$. In a neighborhood of $E=E_0$, the position of the pole in
the plane $\lambda$ will be given by $\lambda=\lambda(E)\equiv\alpha(E)+1/2$, where the function $\alpha(E)$ is
defined as a solution to the equation $f(\alpha(E)+1/2,-k)=0$. But, since $f(\lambda,-k)$ is locally analytic in
the variables $\lambda$ and $k$, with $\partial f/\partial\lambda=\partial f/\partial\alpha\neq 0$, it follows
that $\alpha(E)$ will be, in turn, locally analytic in the given neighborhood of $E=E_0$, so that we may write
$\alpha(E)\approx l+\eta_0+i\sigma_0+\alpha_0'(E-E_0)$ where
$\alpha_0'=(d\alpha/dE)_{E=E_0}\equiv\eta_0'+i\sigma_0'$ and $\sigma_0',\eta_0'$ real numbers. Therefore, in the
neighborhood of $E=E_0$, due to $|\eta_0'|\ll 1, \sigma_0\ll 1$, we have $(l+\alpha+1)\approx 2l+1$ and
$l-\alpha\approx -\alpha_0'[E-E_0+(\eta_0+i\sigma_0)/\alpha_0']=-\alpha_0'(E-E_0+\Delta E+i\Gamma/2)$ where
$\Delta E=(\eta_0\eta_0'+\sigma_0\sigma_0')/|\alpha_0'|^2$ and
$\Gamma=2(\sigma_0\eta_0'-\eta_0\sigma_0')/|\alpha_0'|^2$, so that (155) assumes the following expression
\begin{equation}a_l^R(k)\approx -\frac{1}{\pi(2l+1)}\frac{\beta/\alpha_0'}{E-E_0+\Delta
E+i\Gamma/2}.\end{equation}Under the hypotheses $|\eta_0|\ll 1$ and $\sigma_0\ll 1$, we also have $\Delta E\ll
E_0$ and $\Gamma\ll E_0$, whence, in these conditions, the $l$-th partial wave basically will coincide with
$a_l^R$, so that the contribution of the $l$-th wave, given by
\begin{equation}\sigma_l(k)=4\pi(2l+1)|a_l(k)|^2,\end{equation}to the total scattering cross section, will
be\begin{equation}\sigma_l(k)\approx\frac{4|\beta/\alpha_0'|^2/[(2l+1)\pi]}{(E-E_0+\Delta
E)^2+\Gamma^2/4}\end{equation}which is a formula of Breit-Wigner type, showing that, under the above hypotheses,
we have a resonance, in the wave $l$, for an energy $E=E_0-\Delta E$, with width $\Gamma$. The number $\Delta E$
represents the small variation between the real energy $E_0$, with which the Regge pole goes through nearest
possible to the physical value $l+1/2$ in the plane of $\lambda$, and the real part $E_0-\Delta E$ of the energy
with which the resonance appears in the plane of $E$, for $l$ having a physical value.

Now, as, in the $E$ plane, the imaginary part - proportional to the width $\Gamma$ - of the pole associated with
a resonance, has a well-determined physical meaning related to the mean lifetime $\tau=1/\Gamma$, likewise it is
possible to give, in the $\lambda$ plane or plane of the angular momentum, a direct physical meaning also to
$\sigma_0$, i.e. the imaginary part of the pole associated with the same resonance, as follows. If one imagines,
from a classical standpoint, a resonance as a (semi-bound) metastable system in which the related constituents
rotate around each other for a time $\tau$, then $\sigma_0$ is related with the angular lifetime of such a
resonance. Indeed, if $\lambda=\lambda(E)$ is a value of $\lambda$ for which a Regge pole appears, then we have
\begin{equation}f(\lambda(E),-k)=0\end{equation}and the regular solution $\phi(\lambda(E),k(E),r)$ undergoes the
asymptotic behavior given by (103), that, from (159), reduces to
\begin{equation}\phi\underset{r\rightarrow\infty}{\sim}\frac{1}{2ik}f(\lambda(E),k)
e^{i\vec{k}\cdot\vec{r}}.\end{equation}If $\Im k>0$, we consider the following equations
\begin{equation}\begin{split}&\phi''+\Big(E-\frac{\lambda^2(E)-1/4}{r^2}-U(r)\Big)\phi=0\\&
\dot{\phi}''+\Big(E-\frac{\lambda^2(E)-1/4}{r^2}-U(r)\Big)\dot{\phi}=-\phi+
\frac{d\lambda^2(E)}{dE}\frac{\phi}{dE}\end{split}\end{equation}where $\dot{\phi}=\partial\phi/\partial E$. From
them, it follows that
\begin{equation}\frac{d\lambda^2}{dE}=\frac{\displaystyle\int_0^{\infty}\phi^2dr}
{\displaystyle\int_0^{\infty}\frac{\phi^2}{r^2}dr}\end{equation}so, when $E$ and $\lambda$ are quasi-real
numbers\footnote{See, for instance (Kontolatou 1993) and (Goodman \& Hawkins 2014, Definition 2.2).}, then also
$\phi^2$ will be a quasi-real and non-negative function, hence also $d\lambda^2/dE$ will be a quasi-real and
non-negative function as well, to be precise equal to $R^2$ if $R$ is the classical radius of the orbit of the
system, so that, if $\alpha'=d\alpha/dE=d\lambda/dE=(1/2\lambda)(d\lambda^2/dE)$, then we have $\alpha'\approx
R^2/2\lambda$. Furthermore, $\sigma_0$ is linked to the width $\Gamma$ of the resonance by the approximate
relation $\Gamma/2\approx\sigma_0/\alpha_0'$, so that we have $\sigma_0\approx R^2\Gamma/4\lambda$ that, in
natural units, reduces to $\sigma_0\approx R\Gamma/2v$ because the angular momentum approximately is
$\lambda\approx (1/2)vR$. Therefore, if $\Delta t$ is the revolution period, then we have $\Delta t=2\pi R/v$,
and being $\tau=1/\Gamma$, we have too $\sigma_0\approx (1/4\pi)(\Delta t/\tau)$, so that $\sigma_0/2$
represents the angular mean lifetime $\Delta\theta$ of the system, that is $1/2\sigma_0\approx 2\pi(\tau/\Delta
t)\equiv\Delta\theta$, by which it follows that the lower the imaginary part of the Regge pole is, the greater
$\Delta\theta$ will be. Accordingly, we may conclude that a Regge pole, for
$\lambda=\alpha(E)+1/2=\alpha_l(E)+1/2$, describes, at varying the energy $E$, a trajectory in the plane
$\lambda$; each time that a \it Regge trajectory $\alpha_l=\alpha_l(E)$ \rm goes through very near to an integer
number $l$, then a resonance will appear. Thus, the function $\alpha_l(E)$ itself may give rise to many
resonances, which may also be assembled into families, each of which associated with a single representing
trajectory $\alpha=\alpha_n(E)$, this last idea having given rise to to many other applications and
phenomenological speculations in the field of high and low energy physics. Therefore, to sum up following (Eden
et al. 1966, Chapter 3) and (Mackey 1978, Section 21), bound states of Schr\"{o}dinger equation for a
spherically symmetric potential fall into families characterized by increasing angular momentum and decreasing
binding energy, such a family appearing as a sequence of poles occurring in the successive partial wave
amplitudes $a_l(s)\ \ l\in\mathbb{N}_0$, at increasing values of $s$. As already mentioned above, following a
previous analytic continuation technique employed by S. Mandelstam in 1958 in the case of $f(k,\theta)$, the
theory of complex angular momentum was introduced by Tullio Regge, in 1959, thanks to which it was possible to
show that one could extend $S_l(k)$ in a natural way as to be defined for all positive real $l$, and that
therefore $l\rightarrow S_l(k)$ could be regarded as a boundary value of an analytic function of two complex
variables, the famous Regge poles being the poles of $l\rightarrow S_l(k)$, with respect to $l$ and for fixed
real values of $k$, located in the half-plane $\Re l>-1/2$ where $S_l(k)$ is meromorphic with respect to $l$. To
be precise, the partial scattering amplitude $a_l(k)$ undergoes an \it analytic interpolation \rm from integer
values\footnote{Because such a function is initially defined, in the variable $l$, only for an infinite discrete
set of values, that is $\mathbb{N}_0$, so that it is more correct to speak of an analytic interpolation rather
than an analytic continuation.} to complex values that, for a certain class of Yukawian potentials, is uniquely
determined by the asymptotic behavior of $\Im l\rightarrow\infty$, thanks to Carlson's theorem. At the same
time, with this analytic continuation of $S_l(k)$, considering $s=k^2$ with its negative and positive
determinations, we may also analytically extend $f(s,t)$ by means of a sum of partial scattering amplitudes
$a_l(k)$ via the Watson-Sommerfeld transformation as given by (151) and the relation (64) which links together
scattering phases (i.e., $S_l(k)$) with partial scattering amplitudes (i.e., $a_l(k)$). Therefore, the precise
connection between asymptotic behavior and individual angular momentum singularities was usually achieved by
Regge's method via Watson-Sommerfeld transform (see (Chew 1966, Chapter 8, Section 8-2.)). Varying
$k\in\mathbb{R}$, we obtain a family of curves in the $l$ plane, say $S_l(k_n)$, hence a family of poles, say
$\alpha_l(k_n)$, which therefore change with $k$, so obtaining a discrete family of functions (Regge
trajectories) $k\rightarrow\alpha_l(k)$ at varying of $k$, where $\alpha_l(k)$ is a pole of $l\rightarrow
S_l(k)$ for each $k$ fixed, which identifies its relative position on the given Regge trajectory. For
$s\in\mathbb{R}^-$, $\alpha_l(k)$ is a real increasing function with respect to $k$; then, whenever
$\alpha_l(k)$ passes through a positive integral value $l_0$, then the corresponding value of $k$ is a pole of
$S_{l_0}(k)$ corresponding to a bound state with negative energy; for $k\in\mathbb{R}^+$, $\alpha_l(s)$ is
instead a complex function, with positive imaginary part, so that when $k$ passes from negative values to
positive ones, the Regge trajectory leaves the real axis and enters into the upper half-plane $\Im l>0$ of the
complex $l$ plane. Therefore, the bound states of the dynamical system lie on Regge pole trajectories, several
times on each trajectory, so that the bound states may be grouped into families, all members of the same family
lying on a single Regge trajectory. Again, following (Eden et al. 1966, Chapter 3), bound states of
Schr\"{o}dinger equation for a spherically symmetric potential fall into families characterized by increasing
angular momentum and decreasing binding energy, such a family appearing as a sequence of poles occurring in the
successive partial wave amplitudes $a_l(s)\ \ l\in\mathbb{N}_0$, at increasing values of $s$. The Regge theory
pictures this sequence as due to the presence of a single pole whose position varies continuously with $l$ and
which is relevant to the physics of bound states only when $l$ takes non-negative integral values. This idea has
given meaning by the construction of an interpolating amplitude $a_l(s)$, defined for non-integral, and indeed
complex, values of $l$, which coincides with the physical amplitudes $a_l(k)$ when $l\in\mathbb{N}_0$. This
function $a_l(s)$ is an analytic function of its arguments except for certain singularises. Among these
singularities will be a pole (Regge pole) corresponding to each of the bound state families, the location, or
trajectory, of such a pole being given by the equation $l=\alpha(s)$, the bound state energies corresponding to
values of $s$ which make $l$ take the values $0,1,2,...$. One can deal similarly with unstable bound states and
resonances: indeed, given any pole $k_0$ of $\alpha_{l_0}(k)$ in the complex $k$ plane, we may obtain a nearby
pole in the complex $k$ plane for each value of $l$ near $l_0$. Thus, we obtain a two-dimensional submanifold of
the direct product of the complex $l$ plane with the complex $k$ plane, having the property that each pair
$(l,k)$ in this submanifold is a pair such that $k$ is a pole of $\alpha_l(k)$. These submanifolds will be
finite or countable in number, and we may group together resonances when they lie on the same one, speaking of
\it Regge recurrences \rm of a given bound state or resonance. To be precise, if there exist positive values of
$k$ in which $\alpha_l(k)$ has an integer real part and a very small imaginary part, say
$\alpha(k)=L+i\alpha'(k)$ with $0<\alpha'(k)\ll 1$, then a resonance with angular momentum $L$ and energy $k$,
does appear. The greater the value of $k$, the more the Regge trajectories depart out from the real axis,
towards the plane $\Re l>0$, which, for attractive potentials, form clusters or families of bound states and
resonances.

Therefore\marginpar{\tiny\it Duality}, to summarize, the connection between singularities of the scattering
amplitudes, considered as a complex function analytically prolonged with respect to the energy or the angular
momentum, and the presence of stable (bound states) or unstable (resonances) particles, played a fundamental
role in those theoretical approaches of elementary particle physics in which a central role is given by the
scattering amplitudes and their analytical properties; moreover, the phenomenology of the 1950s high-energy
physics plainly recalled the attention on the importance of singularities of $S$ matrix theory with their
physical meaning, pointing out on the notable contribution of bound states and resonances till to suppose that
non-resonant contributions could be described by means of suitable averages of resonances. This last idea is the
conceptual key of the so-called \it duality, \rm which, in turn, was at the basis of the \it string theory, \rm
whose early origins may be retraced in the late 1960s works of Gabriele Veneziano, and whose main study's object
is the so-called \it relativistic string, \rm a concept generalizing that of particle associated to a given
quantum field, understanding this as represented by the vibrational modes of certain geometrical structures,
having a well-defined dimension, so providing a new theoretical framework for the strong interactions whose
phenomenology of the time shown a prevalence of either diffractive phenomena and resonances which, therefore,
entailed the consideration of a mathematical structure generating infinite states, like a vibrating string.
Following (Marchesini et al. 1976), at the beginnings, attempts to apply quantum field theory for treating
strong interactions were tempted, going beyond the initial and limited Yukawian work on the mesonic field. The
EM analysis of the structure of hadrons has moreover identified a complex structure formed by more elementary
components, called \it partons, \rm to which it will be allowed to apply possibly the ordinary quantum theory of
fields. Therefore, from this standpoint, hadrons should be seen as bound states of partons, while the collisions
between hadrons should be considered as collisions between bound states. Nevertheless, the resulting theoretical
framework presented difficult formal problems like the one regarding the confinement of partons within an
hadron, so that a necessary way to follow for overcoming this type of problems consisted in finding
non-conventional field theories or even completely new formal sights, like the one which revolves just around
$S$ matrix theory, of which one of the most promising formal model was that based on the above notion of duality
as meant by Veneziano. If we consider a general exchange process\footnote{It is a process in which, for
instance, a condition of the following type holds, i.e., the quantum numbers of $a$ and $a'$ are different
between them; this is the case which happens, for example, when the exchange of quarks and gluons between the
hadrons is involved (see (Collins \& Martin 1984, Chapter 7)).} of the type $a+b\rightarrow a'+b'$, and assume
to be valid the hypothesis of duality, then the related scattering amplitude may be given by the sum of two main
contributions: a low-energy contribution given by the Breit-Wigner formula (20), say $a_l^{BW}$, related to a
decay of resonances of the type $a+b\rightarrow(R)\rightarrow a'+b'$, and an high-energy contribution given by
the Regge behavior according to (153), say $a_l^R\approx(k^2)^{\alpha}$, so that we should have $a_l\approx
a_l^{BW}+a_l^R$. The phenomenological analysis of such a question has entailed that, at least in the average
approximation, the related imaginary parts should be equal amongst them, that is $\Im a_l\cong\Im a_l^R\cong\Im
a_l^{BW}$, so that, roughly, the Regge behavior and the formation of resonances basically are dual descriptions
of the same phenomenon. Following (Martin \& Collins 1984, Chapter 7), one of the main characteristic properties
of hadron interactions is that the scattering amplitude may be given as a sum of resonance contributions of the
type $a_l\approx\sum_{res} a_l^{BW}$ with $a_l^{BW}$ given by the Breit-Wigner formula. In the limit of small
resonance width $\Gamma$, we may also write this last sum as a sum of resonance contributions of mass $m_i$
along the direct channel as follows
\begin{equation}a_l(s,t)=\sum_i\frac{1}{s-m_i^2}c_i(t)\end{equation}where $s=E^2=k$ and $t$ is defined as above
and proportional to the square of the four-momentum. But, in the limit $\Gamma\rightarrow\infty$, from what has
been said above, we have that the scattering amplitude may be also written as a sum of Regge contributions as
follows $a_l\approx\sum_{res}a_l^R$, so that, along the crossed channel, we have too
\begin{equation}a_l(s,t)=\sum_i\frac{1}{t-m_i^2}c_i(s),\end{equation}so that duality implies that the scattering
amplitude may be equivalently written both as a sum of resonances related to the direct channel (equation (163))
and as a sum of resonances related to the crossed channel (equation (164)), from which it follows another of the
most characteristic properties of hadron interactions, that is to say the crossing symmetry in the variables $s$
and $t$ (\it duality\rm), so that each of these two formulas expresses a single equivalent contribution to the
scattering amplitude, this last property having been elegantly expressed by the mathematical properties of
following formula for the scattering amplitude
\begin{equation}a_l(s,t)=\frac{\Gamma(1-\alpha(s))\Gamma(1-\alpha(t))}{\Gamma(1-\alpha(s)-\alpha(t)},\end{equation}
proposed by G. Veneziano in 1968, where $\alpha(x)=\alpha_0+\alpha'x$ is a linear Regge's trajectory, $\Gamma$
is the Euler gamma function with poles in $\mathbb{N}_0$ such that $\Gamma(z)\approx z^z$. The symmetries of
(165) take into account the duality principle of above; furthermore, it has poles of resonance type for
$s=m_i^2(i-\alpha_0)/\alpha'$, with highest spin $\alpha_i(m_i^2)$, that, for great values of $s$, gives rise to
the Regge's asymptotic behavior $a_l(s,t)\approx s^{\alpha(t)}$. The interest of (163) and (164) is that, the
duality concept underlying them, may be extended to the case of scattering amplitudes related to an arbitrary
number of particles when one is able to identify the right variables generalizing $s$ and $t$, so that such a
duality hypothesis (with $\Gamma\rightarrow 0$) determines the form of every hadronic scattering amplitude, so
casting the bases of a theoretical framework of hadronic interactions. Nevertheless, not always the basilar
condition $\Gamma\rightarrow 0$ has a reality meaning because, in general, $\Gamma$ grows together with the
resonance mass.

To this point, it is maybe better to follow, almost verbatim, (De Alfaro et al. 1973, Chapter 9), where a new
approach to strong interactions is presented just making use of the notion of duality. In (De Alfaro et al.
1973, Chapter 9, Section 1), the authors point out that in the recent theoretical development classified as
'duality' the substitution rule, or 'crossing', plays a fundamental role. Roughly speaking, in a typical process
\begin{equation}A+B\rightarrow C+D,\end{equation}the same scattering amplitude should describe the process (166),
together with the 'crossed' processes
\begin{equation}A+\bar{C}\rightarrow\bar{B}+D,\qquad A+\bar{D}\rightarrow\bar{B}+C.\end{equation}In the case of
amplitudes with a larger number of external lines, then the number of processes which should be simultaneously
described increases very rapidly with the number of external particles. It should be emphasized that the very
strong predictive power of crossing is present only if one makes further assumptions amongst which analiticity
and 'good' asymptotic behavior. Indeed, since, for example, the three physical regions -- i.e. the allowed
ranges for the kinematical variables -- of the processes (166-67) are separated and disconnected, we should
carry out three independent descriptions of these three processes. Instead, the addition of analyticity
assumptions allows, for instance, a continuation from one physical region to another, so enabling us to obtain
strong constraints among these three different descriptions. In fact, it is very probable that an approximate
amplitude which is perfectly reasonable for reaction (166) proves to be roughly inadequate for the 'crossed'
channels given by the reactions (167). So, the addition of the requirement of 'good' asymptotic properties, like
those provided by Regge prescriptions, has the consequence that any description which is complete for one of the
three channels, should also be complete for the other two. Let us consider, for example, an amplitude satisfying
the dispersion relation\begin{equation}A(s,t)\doteq\frac{1}{\pi}\int\frac{\Im
A(s',t)}{s'-s}ds',\end{equation}where it is assumed that the dispersion relation converges for a sufficiently
spacelike value of $t$, and possible singularities in the third variable $\bar{s}$ are, for simplicity,
neglected. The latter equation (168) shows that a knowledge of the discontinuity of $A(s,t)$, related to those
physical processes in which $s$ is the energy, leads, in principle, to a complete knowledge of $A(s,t)$ and also
of the analytic properties in the variable $t$. For instance, the possible poles in $t$, related to possible
particles exchanged in the $t$ channel, should not be added in an \it ad hoc \rm manner to the right-hand side
of equation (168), but should come from the divergence of the dispersion integral in the variable $s'$. In this
regard, it is well-known that the \it Regge's asymptotic law\begin{equation}\Im
A(s,t)\underset{s\rightarrow\infty}{\sim}\beta(t)s^{\alpha(t)}\end{equation}\rm gives rise just to this type of
behavior, so we see a direct relation between $s$- and $t$-channel descriptions of scattering. A more precise
and very convenient form of expressing such interrelations is given by the so-called \it finite energy sum rules
\rm (FESR) of the form\begin{equation}\int_0^Ls^n\Im
A(s,t)ds=\beta(t)\frac{L^{\alpha(t)+n+1}}{\alpha(t)+n+1}\end{equation}which gives a direct relation between the
$s$-channel description of scattering (that can be inserted into the left-hand side of this equation) and the
$t$-channel Regge poles appearing in the right-hand side. In particular, if we wish to evaluate the left-hand
side of equation (170), inserting only the contribution of the resonance exchanged in the $s$ channel, then the
relation (170) leads to a very important duality property which, broadly speaking, can stated in the following form:\\

\it<<the $t$-channel Regge-pole description of the process corresponds to an appropriate average of the
amplitude computed in terms of the $s$-channel resonances>>.\\\\\rm The duality standpoint has very important
implications in the phenomenological analysis of data: indeed, it tells us that 'interference' models based on
the coherent sum of Regge plus resonance contributions should be discarded. From the theoretical point of view,
the duality relation between $s$-channel resonances and $t$-channel Regge poles, strongly suggests the
possibility of setting up a \it duality bootstrap. \rm Indeed, the $t$-channel Regge pole interpolates a family
of $t$-channel resonances, so that, if we are working in a completely crossing symmetric problem (such as, for
example, $\pi\omega\rightarrow\pi\pi$), then we arrive at a large number of self-consistency constraints for
resonances exchanged in the process. A first semi-phenomenological treatment shows that adjustments can be made
with straight line trajectories and when an infinite number of parallel daughter trajectories, namely
$\alpha_n(s)=\alpha(s)-n$, are also present. The most direct and explicit way of carrying out the
duality-resonance bootstrap leads directly to the Veneziano approach, which was inspired by the use of the
phenomenological approach based upon FESR and duality bootstrap programme applied to the analysis of the process
$\pi\omega\rightarrow\pi\pi$, so giving rise to first dual-resonance amplitude satisfying crossing and Regge
behavior. Veneziano approach is therefore based on the idea that a complete, although approximate, description
of scattering can be obtained in terms of $s$-channel resonances only. If we insert in the dispersion relation
(168) the resonance contributions coming from the particles contained in the trajectory $\alpha(s)$ and in its
daughters, we obtain
\begin{equation}A(s,t)=\sum_n\frac{c_n(t)}{\alpha(s)-n}\end{equation}where $c_n(t)$ is a polynomial of order $n$
in $t$. The crucial point is that the representation (171) is complete and should contain (in an implicit
manner) the poles in the crossed variable $t$. Those poles should arise from the possible divergence of the sum
on the right-hand side of equation (171), which should thus contain an infinite number of terms. We see that one
can hope to achieve a purely resonance bootstrap only in the case of infinitely-rising trajectories. A precise
mathematical solution to duality constraints, namely
\begin{equation}A(s,t)=\sum_{n=0}^{\infty}\frac{c_n(t)}{\alpha(s)-n}=\sum_{n=0}^{\infty}\frac{c_n(s)}{\alpha(t)-n},\end{equation}
has been given by the now famous beta function amplitude of Veneziano, i.e.
\begin{equation}A(s,t)=\frac{\Gamma(-\alpha(s))\Gamma(-\alpha(t))}{\Gamma(-\alpha(s)-\alpha(t)}.\end{equation}It
is easy to show that such an amplitude satisfies the bootstrap requirement given by equation (172) and,
moreover, is the Regge-behaved both in the $s$- and $t$-channels. We can furthermore see that the Veneziano
model is physically acceptable if all trajectories are straight lines (corresponding to infinitely narrow
resonances) with universal slope. If one moreover compares this result with Regge asymptotism, it is therefore
found that elementary crossing allows only one kind of trajectory, the straight line, while deviations from the
straight line will come together with more complicated types of singularities like Regge cuts. Starting from
Veneziano amplitude, it has therefore been possible to develop a general dual-resonance model, which enabled to
construct, in a consistent and systematic way, dual amplitudes for any inelastic process. For example, a
formalism based on harmonic oscillator operators plays a fundamental role in this theoretical approach. But, for
a more general and deeper view of duality theory of strong interactions, we refer to (De Alfaro et al. 1973,
Chapter 9).

\newpage\subsection*{2. On some historical aspects of 1959 Tullio Regge paper} To sum up, following
(Collins \& Squires 1968, Introduction), although the generalization of angular momentum, say $l$, to a complex
variable, and the subsequent transformation of the partial-wave series to an integral in the $l$-plane, was
therefore introduced by Watson in 1918 and revised by Sommerfeld in 1949, it was not until the pioneering work
of Regge in 1959, within $S$ matrix theory framework, that the possible usefulness of the idea in particle
physics was realized. Regge showed that, for a wide class of potentials, the only singularities of the
non-relativistic scattering amplitude in the $l$-plane were poles which move with energy, i.e. poles at
positions $l=\alpha(s)$. These are now called \it Regge poles, \rm and $\alpha(s)$ is called a \it Regge
trajectory. \rm The method was used by Regge to close an important (if not physically very interesting) gap in
previous proofs of the Mandelstam representation for potential scattering. However, the fact that the Regge
trajectory correspond to physical particles (or resonances) when $\alpha(s)$ equals an integer for positive $s$,
and also determine the high energy behavior of the cross-channel amplitudes, claimed the attention of elementary
particle theorists, and since the early 1960s, the subject had a vigorous, controversial and exiting history. In
(Frautschi 1963, Chapter X), it is also referred that Regge transformed the scattering amplitude to a new
representation involving complex $l$ just by means of the so-called Watson-Sommerfeld transformation. This
method, says Frautschi, had a history stretching over several decades, and was used to study rainbows,
propagation of radio waves around the Earth, and scattering from various potentials. Regge's original
contribution lay in understanding the special features of complex angular momenta for scattering from
superpositions of Yukawa potentials (i.e., the type of potential deemed to be relevant for relativistic
scattering) and calling these features to the attention of high-energy physicists. On the basis of what told by
E. Guth, Frautschi says that a related transformation similar to that handled by Regge, was previously used by
Poincaré and Nicholson in the early 1910s in connection with the bending of electromagnetic waves by a sphere,
hence explicitly introduced by Watson in 1918 and later resurrected by Sommerfeld in 1949. In this regard,
Frautschi also quotes a previous related work of Nicholson dating back to 1907, and not cited by other
references. In any event, already Poincaré, Nicholson and Watson were aware of the existence of certain poles in
the scattering amplitude of acoustic and radio wave propagation problems, then systematically treated by
Sommerfeld who provided a first exhaustive theoretical framework mathematically similar to the one then
established by Regge in dealing with De Broglie's waves, whose pioneering works of the late 1950s opening, for
the first time, the use of such formal techniques in elementary particle physics, clearly underscoring
differences and possible analogies between the acoustic-electromagnetic case and the quantum one.

In the monograph (De Alfaro \& Regge 1965, Chapter 1, Section 1.2), the authors give a compendious historical
outline of potential theory. In particular, they point out what usefulness has their approach in reconstructing
the scattering amplitude via Born approximation method, which uses techniques based on the introduction of a
unique analytic interpolation of partial scattering amplitudes for complex values of the angular momentum,
following 1959 Regge work. Then, in the next Chapter 2, the authors summarize the main mathematical tools used
throughout the text, amongst which is a Poincaré's theorem already quoted in the previous section 1, that they
say to play a key role in potential scattering. This theorem concerns with general linear second order
homogeneous differential equations of the form
\begin{equation}\frac{d^2f(x)}{dx^2}+p(x)\frac{df(x)}{dx}+q(x)f(x)=0\end{equation}which, following (Whittaker \&
Watson 1927, Chapter X, Section 10.1), can be written in the form
\begin{equation}\frac{d^2\varphi(x)}{dx^2}+J(x)\varphi(x)=0\end{equation}where
\begin{equation}f(x)=\varphi(x)\exp\Big(-\frac{1}{2}\int_b^xp(y)dy\Big)\end{equation}and
\begin{equation}J(x)=q(x)-\frac{1}{2}\frac{dp(x)}{dx}-\frac{1}{4}p^2(x).\end{equation}For example, Schr\"{o}dinger
equation for partial waves may be written in the form (175). The above Poincaré theorem (see (Poincaré 1884a,b))
has to do with equations of the form (175) depending on a parameter $\eta$ through a function $J$ of the type
$J(x,\eta)$, supposed to be an analytic entire function of $\eta$. Take now a solution $\psi(x)$ defined by a
boundary condition independent on $\eta$ in an ordinary point $P(x=c)$. The theorem states that $\psi(x,\eta)$
for fixed $x$, as a function of $\eta$, is also an entire function. The condition that $P$ should be an ordinary
point, can be relaxed provided the boundary conditions are still $\eta$-independent. Hence, De Alfaro and Regge
say that many of the theorems which will be established later in the text, are really generalizations of
Poincaré's theorem, this fact witnessing the importance of such a result in the formal framework of potential
scattering. For instance, we have used Poincaré's theorem to prove analyticity properties of a regular solution
to (71). In (Regge 1959), it was considered, for the first time, the possibility of introducing the angular
momentum as a complex variable, showing the convenience of this procedure in proving that Mandelstam
representation for potential scattering holds. Regge starts with the consideration of the Schr\"{o}dinger
equation wrote in the form
\begin{equation}\psi''-\Big(\frac{\lambda^2-(1/4)}{x^2}-1+U(x)\Big)\psi=0\end{equation} having chosen
dimensionless variables, by putting $x=kr$, where $r$ is the distance from the origin, and $k$ the (fixed) wave
number. Here $\lambda$ is a generalized complex orbital momentum, and when it assumes positive half-integer
values (hereafter referred to as the physical values), then we shall write $\lambda=j+1/2$. The equation (178)
is even in $\lambda$, and further restrictions are assumed for $U(x)$ (see (72)). Then Regge considers a
solution to (178), say $F(\lambda,\eta,x)$, whose parameter $\eta$ may vary in such a manner to comprehend other
solutions to (178) (see (Regge 1959, (1.3) and (1.4)); moreover, a slight generalized form of the above
mentioned theorem of Poincaré, states that $F(\lambda,\eta,x)$, for fixed $\eta$ and $x$, is an entire function
of $\lambda$. At the point 6. of his paper, Regge establishes some results in the field of dispersion relations.
As is well-known, these relations are statements of analyticity of the scattering amplitude as a function of the
energy and of the transmitted momentum. Although the energy is kept fixed, it is still possible to derive, for
special classes of potentials, enough properties as to guarantee for the existence of such relations. As we have
seen in the previous section, analyticity in $\cos\theta$ is known to subsist within the Lehmann ellipse (138),
whose corresponding representation is unable for studying the behavior of the scattering amplitude letting
$t\rightarrow\infty$ or equivalently as $\cos\theta\rightarrow\infty$. But Regge cleverly and ably overcomes
this formal problem transforming the series of the partial wave expansion of the scattering amplitude given by
(134) (i.e., the expansion (6.3) of (Regge 1959)), into an integral over the complex plane of the variable
$\lambda=l+1/2$ in such a manner to have an analytic interpolation of the scattering amplitude over an open
domain of the $\cos\theta$-plane, so obtaining the integral (147) (i.e., the integral (6.4) of (Regge 1959) or
the integral (9.19) of (De Alfaro \& Regge 1965, Chapter 9, Section 9.3)), with an artifice that Regge says as
due to G.N. Watson and used by A. Sommerfeld in some wave propagation problems, but without obviously giving
other historical informations in this regard. In (De Alfaro \& Regge 1965, Chapter 13, Section 13.2), the
authors say that, strangely enough, formulas of the type (147) for general singular potential were first
historically treated in connection with the propagation of waves all around the Earth, where the potential
$U(r)$ was a truly hard-core potential, i.e. $U(r)=\infty$ for $r<R$, where $R$ stood for the radius of the
Earth supposed to have a spherical form besides to be perfectly conducting. The resulting pole expansion was
seen to converge very rapidly for large $k$ but there is no indication that this result may be extended to cases
of significance in nuclear physics, as deemed by De Alfaro and Regge in the 1960s.

Because of the importance of this last crucial formal passage with respect to the whole Regge's work achieved in
(Regge 1959) -- whose primary aim was to prove the double dispersion relation in potential scattering -- as well
as for the notable role played by it in the next developments of high-energy physics, we now briefly devote a
lot of time to deepen just these latter historical aspects of such a 1959 Regge's seminal paper. Following
(Newton 1964, Chapter 1), the mathematical analysis of the results of scattering experiments has lately acquired
a new tool based on the replacement of an infinite series of discrete terms by a contour integral of an analytic
function introduced by Poincaré and Nicholson, and first applied in its present form in electromagnetic theory
by Watson in 1918 and, after its resurrection, in electromagnetic wave problems, by Sommerfeld. Also in (De
Alfaro \& Regge 1965, Chapter 9, Section 9.3) some further historical clarifications are included. Indeed, the
authors recall that the form of the scattering amplitude (147) is essentially derived by the work of Poincaré in
diffraction of electromagnetic waves and exposed in (Poincarè 1910) which, nevertheless that, is not quoted in
the original Regge's paper of 1959. In (Poincaré 1910), the author considers applications of this his theorem
quoted above to certain differential equations of the type (178) (like, for example, the equation (2) of Section
12 or the equation (1) of Section 11), as well as Legendre polynomial expansions similar to (134) (like, for
example, the series (2) of Section 6) and integral expression similar to (147) (like, for example, the
expressions (3) or (13) of Section 13 or the expression (3) of Section 15; see also Sections 17 and 19). As his
usual style, Poincaré does not quote any reference, except a mention to Max Abraham (1875-1922), at page 170,
and to Arnold Sommerfeld (1868-1951), at page 171. Afterwards, De Alfaro and Regge quote too the works of John
William Nicholson (1881-1955), namely (Nicholson 1910a,b; 1911), which are centered around mathematical problems
inherent in diffraction phenomena of electromagnetic waves sent by an Hertzian oscillator round the surface of
the Earth considered as a sphere of perfect conductivity, taking into account the previous work made by Poincaré
on this subject in which the electromagnetic forces involved are expressed as an integral of Fredholm's type
which will be then expanded into a series of zonal harmonics. In Nicholson's works, the main formal aspect of
his treatment of the related physical problem is just centered on the technique of replacing a series with an
integral extended to certain infinite regions (see, above all, (Nicholson 1910b) where integral expression of
the magnetic forces are deduced from their series expansions) whose values are then determined by means of the
calculus of residues. On the other hand, as early as 1900s, in many investigations of applied mathematics, it is
necessary to sum a series into each term of which a large parameter enters. For example, many problems of
diffraction of waves by large obstacles depend for their ultimate solution upon the sum of a series of harmonic
terms in which complicated derivates of Bessel functions of large argument are present. By expressing these
transcendental functions in an appropriate manner, these summations may all be made to depend upon a single
type, with which the paper (Nicholson 1907) proposes to deal, and in which series to be summed are, at a first
approximation, replaced by a suitable integral built up according to a certain new method which will be further
retaken by Nicholson himself in the next papers of 1910s. Then, De Alfaro and Regge refer that such a
mathematical technique was extensively used by George Neville Watson (1886-1965) in (Watson 1918) (see also
(Watson 1944) and references therein), where Poincaré's and Nicholson's works are quoted, together that of H.M.
Macdonald (see (Macdonald 1914)), H.W. March, M. von Rybczynski and A.E.H. Love (see (Love 1915) and references
therein). In any case, ever since the time, about 1902, when G. Marconi first succeeded in sending wireless
signals across the Atlantic ocean, the question of explaining the mechanism of such transmission attracted the
attention amongst mathematicians and physical-mathematicians, many of whom are quoted in (Love 1915), amongst
whom are H.M. Macdonald, J. Zenneck, H.W. March, M. von Rybczynski, L.W. Austin, J.L. Hogan, and W.H. Eccles
(see (Love 1915) and references therein), and this ever since early 1900s. But, just following what is said in
(Love 1915, No. 10), for the infinite series which represents the effect of curvature without resistance,
methods of summation were devised by Macdonald (see (Macdonald 1911)), Poincaré and Nicholson (see their above
quoted works, in addition to (Poincaré 1903)); and again another method was devised by Macdonald (see (Macdonald
1911)). All these methods depend upon a transformation of the series into a definite integral, hence by an
approximate evaluation of the latter integral. Poincaré, in (Poincaré 1910), did not press his method so far as
to tabulate numerical results, but concluded that the expression for the electric force normal to the surface,
at an angular distance $\theta$ from the originating doublet, should contain a factor of the type
$e^{-\Lambda\theta}$. Nicholson, in (Nicholson 1910a,b; 1911), went further in the same direction, obtained a
formula for the magnetic force containing such an exponential as a factor, and deduced definite numerical
results. Macdonald, in (Macdonald 1911), also obtained definite numerical results which cannot be reconciled
with those of Nicholson. The discrepancy was discussed by Nicholson himself, who traced it to an alleged flaw in
the analysis used by Macdonald in (Macdonald 1911); it was also discussed by Macdonald in (Macdonald 1914), who
pointed out a difficulty in the analysis used by Poincaré in (Poincaré 1910) and by Nicholson in (Nicholson
1910a,b; 1911). Fresh numerical results were then deduced by Macdonald in (Macdonald 1914) from a new method of
summing the series, but they did not agree with those previously found by Nicholson, or with those previously
found by Macdonald himself in 1911. Nevertheless, many of the authors and related works quoted in (Love 1915),
which have to do with the above crucial passage from a series to an integral, are not yet mentioned in (De
Alfaro \& Regge 1965).

The 1918 paper of Watson, however, put into reciprocal comparison mainly the works of Nicholson and Poincaré on
the one hand, with the works of Macdonald on the other hand, so reaching to a major clarification either in
physical-mathematics consequences and formal aspects of the original physical problem under examination. Under
advice of B. Van der Pol, who asked him to further analyze the problem with the chief purpose to clarify why
there subsist the various discrepancies found by different authors in dealing with this physical problem, Watson
first highlights, from a further comparison between H.W. March and M. Von Rybczynski works with that of A.E.H.
Love and with those of Poincaré and Nicholson, what follows (see (Watson 1918, No. 2))\\

\it<<The essential advance in this paper is closely connected with the fundamental error of March and Rybczynski
which was pointed out by Love. In dealing with an oscillator on the positive half of the axis of harmonics,
those writers express a Hertzian function by an integral of $P_s(\cos\theta)$, the integration being carried out
with regard to the degree $s$ of the Legendre function; such an integral has a line of singularities along the
line $\theta=\pi$, and is regular along the line $\theta=0$. The fact is that, when harmonics of non-integral
degree are introduced, the appropriate function to use is not $P_s(\cos\theta)$ but $P_s(-\cos\theta)$; this
fundamental point is somewhat obscured by the equation$$P_n(-\cos\theta)=(-1)^nP_n(\cos\theta),$$which holds
between the functions whose degrees are integers. The failure of convergence of an integral involving
$P_s(\cos\theta)$ along the line $\theta=0$ (when an oscillator is placed on the positive half of the axis of
harmonics) is strictly analogous to the failure of convergence of the series $1+z+z^2+...$, all round the circle
$|z|=1$ on account of the singularity of the function $1/(1-z)$ at the point $z=1$. A simple electrostatic
example is afforded by the potential of a unit charge at distance $a$ from the origin. The potential near the
origin is$$V=\frac{1}{a}\sum_{n=0}^{\infty}\Big(\frac{r}{a}\Big)^nP_n(\cos\theta)=\leqno(\star_1)$$
$$=\frac{1}{a}\sum_{n=0}^{\infty}(-1)^n\Big(\frac{r}{a}\Big)^nP_n(\cos\theta)=$$
$$=\frac{1}{2ia}\int\Big(\frac{r}{a}\Big)^sP_s(\cos\theta)\frac{ds}{\sin s\pi}\leqno(\star_2)$$where the contour starts from
$+\infty$ and returns to $+\infty$ after encircling the points $s=0,1,2,...$ which are poles of the integrand.
On swinging round the contour so as to surround the other poles of the integrand, and evaluating the residues,
we find the series for $V$ in descending powers of $r$, valid when $r>a$>>.\\\\\rm Watson, in discussing the
swinging of the contour of integration of $(\star_2)$, makes reference to a paper of Ernest William Barnes
(1874-1953) (see (Barnes 1908)) in which the so-called \it Mellin-Barnes integral \rm is defined, as well as he
quotes a Laplace's formula for $P_s(\cos\theta)$ to prove the convergence of the integral for $|s|$ large (see
(Watson 1944, Chapter 6, Section 6.5)). Therefore, it is more probable that Watson, in achieving the integral
$(\star_2)$, considered what Barnes made in this respect in defining his integral which is, roughly, a contour
integral of products involving gamma function and exponential factors (see also (Barnes 1910)). On the other
hand, recent history of mathematics research tells us that it was Salvatore Pincherle (1853-1936), in his works
on generalized hypergeometric functions, to have early applied the elements of the technique of contour
integration methods to special functions (to be precise, to hypergeometric functions) with the introduction of
first forms of integrals of Mellin-Barnes type. Indeed, following (Mainardi \& Pagnini 2003, Section 1),\\

\it<<In Vol. 1, p. 49 of \rm Higher Transcendental Functions \it of the Bateman Project, we read ''Of all
integrals which contain gamma functions in their integrands the most important ones are the so-called
Mellin-Barnes integrals. Such integrals were first introduced by S. Pincherle, in 1888 \rm [see (Pincherle 1888;
1965)]\it; their theory has been developed in 1910 by H. Mellin \rm [see (Mellin 1910)] \it (where there are
references to earlier work) and they were used for a complete integration of the hypergeometric differential
equation by E.W. Barnes \rm [see (Barnes 1908)]. \it In the classical treatise on Bessel functions by Watson \rm
[see (Watson 1944, Chapter 6, Section 6.5)] \it we read: ''By using integrals of a type introduced by Pincherle
and Mellin, Barnes has obtained representations of Bessel functions which render possible an easy proof of
Kummer's formula \rm[...]. \it Here we point out that the 1888 paper (in Italian) of S. Pincherle on the
Generalized Hypergeometric Functions led him to introduce the afterwards named Mellin-Barnes integral to
represent the solution of a generalized hypergeometric differential equation investigated by Goursat in 1883.
Pincherle's priority was explicitly recognized by Mellin and Barnes themselves, as reported below. In 1907
Barnes \rm [see (Barnes 1907)], \it wrote: ''The idea of employing contour integrals involving gamma functions
of the variable in the subject of integration appears to be due to Pincherle, whose suggestive paper was the
starting point of the investigations of Mellin (1895) though the type of contour and its use can be traced back
to Riemann''. In 1910 Mellin \rm [see (Mellin 1910)], \it devoted a section (§ 10: Proof of Theorems of
Pincherle) to revisit the original work of Pincherle; in particular, he wrote ''Before we are going to prove
this theorem, which is a special case of a more general theorem of Mr. Pincherle, we want to describe more
closely the lines $L$ over which the integration preferably is to be carried out''>>.\\\\\rm Finally,
Sommerfeld, in (Sommerfeld 1949, Appendix II to Chapter V, and Appendix to Chapter VI), retakes the above
mentioned work of Watson in computing the solution $u$ to the wave equation $\Delta u+k^2u=0$, of which a first
form is obtained as a series expansion of the type\footnote{Here, $\zeta_n(x)\doteq\sqrt{\pi
x/2}H^{(2)}_{n+1/2}(x)$ (where $H^{(2)}_{\alpha}(x)$ is the Hankel function of the second kind), is the
so-called \it Riccati-Bessel function, \rm while $\xi_n(x)\doteq\zeta_n(x)+x\zeta'_n(x)$.}$$u=\frac{k}{4\pi
i}\sum_{n=0}^{\infty}(2n+1)P_n(\cos\theta)\frac{\zeta_n(kr)}{\xi_n(ka)}\leqno(\ast)$$whose convergence, however,
is very poor, and limited to the domain given by $a<r<\infty$ and $0\leq\theta\leq\pi$. This series expansion
has been studied either by P. Debye in 1908-09 as well as by P. Frank and R. von Mises in the second volume of
the well-known \it Die Differential- und Integralgleichungen der Mechanik und Physik, \rm edited by P. Frank
with the assistance of H. Faxen, R. F\"{u}rth, Th. von K\^{a}rm\^{a}n, R. von Mises, Fr. Noether, C. W. Oseen,
A. Sommerfeld, and E. Trefftz, Braunschweig, Vieweg, 1927-35, whose second volume is a revised and enlarged
edition of the second part of the celebrated 1910 Riemann-Weber treatise \it Die Differential- und
Integralgleichungen der Mechanik und Physik. \rm Hence, Sommerfeld transforms $(\ast)$ in a complex integral,
and, to this end, on the basis of the relation $P_n(\cos\theta)=(-l)^n P_n(-\cos\theta)$, which is valid for
integral (and only for integral) values of $n$, he first rewrites the series in $(\ast)$ in the following form
$$u=\sum_{n=0}^{\infty}(2n+1)(-1)^nP_n(-\cos\theta)\frac{\zeta_n(kr)}{\xi_n(ka)},\leqno(\ast_1)$$hence, he replaces
$n$ by a complex variable $\nu$ and traces a loop $A$, around the real axis of the $\nu$-plane, in such a manner
it surrounds all the points $\nu=0,1,2,3,...,n,...$ in a clockwise direction. Over this loop, Sommerfeld takes
the integral
$$u=\int\frac{2\nu+1}{2i\sin\nu\pi}P_{\nu}(-\cos\theta)\frac{\zeta_n(kr)}{\xi_n(ka)}d\nu\leqno(\ast_2)$$
which is obtained from the general term in $(\ast_1)$ by interchanging $n$ and $\nu$, suppressing the factor
$(-1)^n$, and appending the denominator $\sin\nu n$. As in (Sommerfeld 1949, Chapter V, Appendix II), $P_{\nu}$
does not stand for the Legendre polynomial, but for the hypergeometric function
$$P_{\nu}(x)=F\Big(-\nu,\nu+1,1,\frac{1-x}{2}\Big)$$which is identical with the Legendre polynomial only for integral
$\nu$. In doing so, Sommerfeld resembles 1888 Pincherle's method. Then, Sommerfeld suitably deforms, in many
ways, the path $A$ taking into account poles and zeros of the integrand of $(\ast_2)$ and applying Cauchy's
residue theorem, hence discussing the related consequences in relation to $(\ast_2)$ until up to obtain the
following expression
$$u=\pi\sum_{\nu=\nu_0,\nu_1,...}\frac{2\nu+1}{2i\sin\nu\pi}P_{\nu}(-\cos\theta)\frac{\zeta_n(kr)}{\eta_n(ka)}\leqno(\ast_3)$$
where, in a neighborhood of the $m$-th pole of the integrand of $(\ast_2)$,
$\eta_{\nu}=(\partial\xi_{\nu}/\partial\nu)_{\nu=\nu_m}$ and $\xi_{\nu}(ka)=(\nu-\nu_m)\eta_{\nu}(ka)$, if one
considers the zeros of $\xi_{\nu}(ka)=0$ for $\nu=\nu_0,\nu_1,...$ as the poles of the integrand of $(\ast_2)$.
Now, except for a sign and for a constant factor, the integral $(\ast_2)$ is identical with the solution
$(\ast_1)$ of the sphere problem. Hence the series $(\ast_3)$ also represents the solution of the sphere
problem, and suppressing the immaterial constant factor, Sommerfeld finally writes
$$u=\sum_{\nu}\frac{2n+1}{\sin\nu\pi}P_n(-\cos\theta)\frac{\zeta_{\nu}(kr)}{\eta_{\nu}(ka)},\leqno(\ast_4)$$so
that the passage from the series $(\ast_1)$, which is summed over integral $n$, to the series $(\ast_4)$, which
is summed over the complex $\nu$, is obtained by forming residues in a complex integral twice. This, finally,
explains why De Alfaro and Regge, in (De Alfaro \& Regge 1965, Chapter 9, Section 9.3) (see also (De Alfaro et
al. 1973, Chapter 1, Section 7)) speak of \it Watson-Sommerfeld transform \rm (in short, \it WS-transform \rm or
\it SW-transform\rm), even if, in this paper, we have tried to deepen what have been the real historical roots
of this transform, identifying these in the previous works of E.W. Barnes and S. Pincherle.

In conclusion, in this last section we have only pointed out some little known historical perspectives which
underlie the introduction of WS-transform, identifying the early roots upon which it relies. As we have seen in
the previous Section 1, after the pioneering use of the WS-transform involving complex angular momenta for
studying resonance phenomena of elementary particles mainly motivated to prove the validity of the Mandelstam
representation for the potential scattering of two spinless particles for a certain class of generalized Yukawa
potentials, it was possible to work out the notions of Regge pole and Regge trajectory, from which Gabriele
Veneziano started his as many pioneering work of 1968 (see (Veneziano 1968)) in which resonance dual models are
introduced, for the first time, in fundamental physics. Veneziano's model basically begins with the proposing of
a quite simple expression for the relativistic scattering amplitude (today known as \it Veneziano amplitude\rm)
that obeys the requirements of Regge asymptotics and crossing symmetry in the case of linearly-rising
trajectories, containing automatically Regge poles in families of parallel trajectories with residue in definite
ratios, which furthermore satisfies the conditions of superconvergence (S. Fubini) and exhibits, in a nice
fashion, the duality between Regge poles and resonances in the scattering amplitude, essentially based on
symmetry properties of Euler beta function (see (Green et al. 1988, Chapter 1, Sections 1.1 and 1.2)). This is
the incipit of a coherent $S$ matrix theory of strong interactions which then will led to resonance dual models
in high-energy particle physics, so, as is well known, marking the rising of string theory. Indeed, following
(Di Vecchia 2008), once that Regge discovered that the Schr\"{o}dinger equation allowed continuation in angular
momentum for complex values, and linked resonances with different spin, one of the basic ideas that led to the
construction of an $S$ matrix was that it should include resonances at low energy and at the same time give
Regge behavior at high energy. But the two contributions of the resonances and of the Regge poles should not be
added because this would imply double counting. This was called \it Dolen-Horn-Schmidt duality, \rm discovered
around 1967, while another idea (of about 1968) that helped in the construction of an $S$ matrix was the
so-called \it planar duality \rm that was visualized by associating to a certain process a duality diagram where
each meson was described by two lines representing the quark and the antiquark. Finally, also the requirement of
crossing symmetry played a very important role, and starting from these ideas Veneziano was finally able to
construct an $S$ matrix for the scattering of four mesons that, at the same time, had an infinite number of zero
width resonances lying on linearly-rising Regge trajectories and Regge behaviour at high energy. In any case,
after the Regge 1959 seminal paper, the WS-transform was extensively used in field theory as, for instance, done
in (Schwarz 1973, Section 1), where, on the basis of an axiomatic scheme (amongst which the assumption that all
mesons and baryons, together with the related poles, lie on a Regge trajectory), it is considered the following
dispersion relation$$A(s,t)=\sum_{n=0}^{\infty}\frac{R_n(t)}{n-\alpha(s)},$$to which the scattering amplitude
must satisfy, together with its WS-transform, namely$$A(s,t)=\frac{1}{2\pi
i}\int_{\mathcal{C}}\tilde{R}(t)\frac{\Gamma(-n)}{n-\alpha(s)}dn$$whose WS-contour $\mathcal{C}$ will be
suitably deformed to avoid the pole at $n=\alpha(s)$. Through a suitable choice of the function $\tilde{R}(t)$,
Schwarz derives Veneziano amplitude just by means of such a method based on WS-transform. Likewise to the paper
(Mandelstam 1974), where, although not explicitly mentioned, the WS-transform plays a fundamental role since the
beginnings. Furthermore, following (Cushing 1990, Chapter 5, Section 5.4), the asymptotic behavior of
$f(k,\theta)$ as $\cos\theta\rightarrow\infty$, was settled for potential scattering by Regge in 1959. Although
a beautiful piece of classical mathematical analysis in its own right, this paper was to have its greatest
impact on the $S$ matrix program for the conjectures to which it would lead. Only (Collins 1977, Chapter 2,
Sections 2.7, 2.9 and 2.10; Chapter 4, Section 4.6; Chapter 7, Section 7.1; Chapter 9, Section 9.3) has devoted
the right attention to the WS-transform, there called \it Mandelstam-Watson-Sommerfeld transform, \rm after the
improvement of the original WS-transform by Mandelstam himself in 1962, recalling as well the links between it
and the Mellin transform; in this regard, see also (Collins \& Squire 1968, Chapter II, Section II.9) and (Omnès
\& Froissart 1963, Chapter 4, Sections 4-2 and 4-3), where interesting historical notes on Sommerfeld original
work have also been inserted. In (Frautschi 1963, Chapter X), as regard Watson-Sommerfeld transformation, it is
then said that this method had a history stretching over several decades and had been used to study rainbows,
propagation of radio waves around the Earth, and scattering from various potentials, the author specifying that
the related transformation was used by Poincaré and Nicholson in 1910 in connection with the bending of
electromagnetic waves by a sphere, the transformation having been introduced in its present form by Watson in
1918 and later resurrected by Sommerfeld. Regge's original contribution lays in understanding the special
features of complex angular momenta for scattering from superpositions of Yukawa potentials -- the type of
potential believed relevant to relativistic scattering -- and calling these features to the attention of
high-energy physicists. In (Frautschi 1963, Chapters X and XIII), besides further and interesting its
applications in high-energy physics, it is also quoted who has later improved and extended WS-transformation.

Finally, in this paper, we have pointed out a single historical aspect underling one the most important works of
20th century physics, that is to say, the 1959 seminal paper of Tullio Regge. Following (Schwarz 1975), Regge's
pioneering work of 1959 provided, amongst other things, a remarkable classification of hadrons by means of the
so-called Regge's trajectories. Furthermore, as has been seen above, this Regge's work has mainly a technically
fashion in which new formal methods and techniques have been introduced, besides to have introduced new
pioneering and fruitful ideas and notions having physical nature. Among these techniques, just the so-called
Watson-Sommerfeld transform, on which we have focussed our historical sight with the main purpose to identify
the early origins and sources upon which rely such a complex transform. In pursuing this aim, we have found
early sources of this formal method in some works dating back the early 1900s and mainly due to Poincaré,
Nicholson, Macdonald, and Barnes, until to reach to descry some prolegomena, or protohistoric origins, of the
WS-method in some previous works of Pincherle of the late 1880s. Because of the notable importance played by
this transform in the formal framework of fundamental physics, we think that this our preliminary historical
work, albeit centered on a single methodological aspect, is not altogether superfluous if nothing else for the
simple fact that such a historical clarifications have not been treated in a whole and deep manner if not fully
neglected. Above all, we have found a very few historical accounts on WS-transform, which we have seen to be a
crucial technique laying out at the early foundational bases of resonance dual models of elementary
particles.\\\\\bf Acknowledgements. \rm This paper had been thought as an enlarged and complete version to
support my presentation to be held at the the XXX-th International Workshop on High Energy Physics \it
<<PARTICLE and ASTROPARTICLE PHYSICS, GRAVITATION and COSMOLOGY: PREDICTIONS, OBSERVATIONS and NEW PROJECTS>>,
\rm which was planned to be held in the \rm which was planned to be held in the National Research Centre
''Kurchatov Institute'' of the Institute for High Energy Physics of Protvino, Russia, in June 23-27, thanks to
the kind invitation by Nykolay Tyurin, Vladimir Petrov and Roman Ryutin, CMS members of CERN, Geneva. My sincere
gratitude to them, even if it wasn't possible, with my regret, to attend to this renowned conference.

\newpage\section*{Minimal Bibliography}
\begin{description}\item Barnes, E.W. (1907), ''The asymptotic expansion of integral functions defined by
generalized hypergeometric series'', \it Proceedings of the London Mathematical Society, \rm 5 (2): 59-116.\item
Barnes, E.W. (1908), ''A New Development of the Theory of the Hypergeometric Functions'', \it Proceedings of the
London Mathematical Society, \rm 6 (2): 141-177. \item Barnes, E.W. (1910), ''A Transformation of the
Generalized Hypergeometric Series'', \it Quarterly Journal of Mathematics, \rm 41: 136-140.\item Calogero, F.
(1967), \it Variable phase approach to potential scattering, \rm New York, NY: Academic Press, Inc. \item Chew,
G.F. (1962), \it $S$-Matrix Theory of Strong Interactions. A Lecture Note Reprint Volume, \rm New York, NY: W.A.
Benjamin, Inc.\item Chew, G.F. (1966), \it The Analytic $S$ Matrix. A Basis for Nuclear Democracy, \rm New York,
NY: W.A. Benjamin, Inc.\item Cini, M. (1977), ''Storia e ideologia delle relazioni di dispersione'', in: Donini,
E., Rossi, A. e Tonietti, T. (a cura di) (1977), \it Matematica e Fisica: struttura e ideologia, \rm Atti del
Convegno \it Aspetti strutturali e ideologici nel rapporto tra scienze fisiche e matematiche, \rm promosso e
organizzato dall'Istituto di Fisica dell'Università di Lecce, 1-5 Luglio 1975, Bari, IT: De Donato editore.\item
Collins, P.D.B. and Squires, E.J. (1968), \it Regge Poles in Particle Physics, \rm Springer Tracts in Modern
Physics, Volume No. 45, Berlin \& Heidelberg, DE: Springer-Verlag.\item Collins, P.D.B. (1977), \it An
Introduction to Regge Theory \& High Energy Physics, \rm Cambridge, UK: Cambridge University Press.\item
Collins, P.D.B. and Martin, A.D. (1984), \it Hadron Interactions, \rm Bristol, UK: Adam Hilger, Ltd.\item
Cushing, J.T. (1986), ''The Importance of Heisenberg's $S$-Matrix Program for the Theoretical High-Energy
Physics of the 1950's'', \it Centaurus. An International Journal of the History of Science and Its Cultural
Aspects, \rm 29 (2): 110-149.\item Cushing, J.T. (1990), \it Theory construction and selection in modern
physics. The $S$ Matrix, \rm Cambridge, UK: Cambridge University Press.\item De Alfaro, V. and Regge, T. (1965),
\it Potential Scattering, \rm Amsterdam, NL: North-Holland Publishing Company.\item De Alfaro, V., Fubini, S.,
Furlan, G. and Rossetti, C. (1973), \it Currents in Hadron Physics, \rm Amsterdam, NL: North-Holland Publishing
Company.\item Di Vecchia, P. (2008), ''The Birth of String Theory'' in: Gasperini, M. and Maharana, J. (Eds.)
(2008), \it String Theory and Fundamental Interactions. Gabriele Veneziano and Theoretical Physics: Historical
and Contemporary Perspectives, \rm Lecture Notes in Physics, No. 737, Heidelberg and Berlin, DE:
Springer-Verlag, pp. 59-118.\item Eden, R.J., Landshoff, P.V., Olive, D.I. and Polkinghorne, J.C. (1966), \it
The Analytic $S$-Matrix, \rm Cambridge, UK: Cambridge University Press. \item Fock, V.A. (1943) ''On the
representation of an arbitrary function by an integral involving Legendre functions with complex index'', \it
Doklady Akademii Nauk, SSSR, \rm 39: 253-256 (in Russian). For an English Translation, see Faddeev, L.D.,
Khalfin, L.A. and Komarov, I.V. (Eds.) (2004), \it V.A. Fock. Selected Works on Quantum Mechanics and Quantum
Field Theory, \rm Boca Raton (FL): Chapman \& Hall/CRC - A CRC Press Company, LLC, 43-1, pp. 494-499.\item
Frautschi, S.C. (1963), \it Regge poles and $S$-matrix theory, \rm New York, NY: W.A. Benjamin, Inc.\item
Goodman, S. and Hawkins, J. (2014), ''Julia Sets on $\mathbb{R}\mathbb{P}^2$ and Dianalytic Dynamics'', \it
Conformal Geometry and Dynamics. An Electronic Journal of the American Mathematical Society, \rm 18:
85-109.\item Gordon, W. (1928), ''\"{U}ber den Sto\ss\ zweier Punktladungen nach der Wellenmechanik'', \it
Zeitschrift f\"{u}r Physics, \rm 48 (3-4), 180-191.\item Green, M.B., Schwarz, J.H. and Witten, E. (1988), \it
Superstring Theory, \rm Volume 1: Introduction, Cambridge, UK: Cambridge University Press.\item Khuri, N.N.
(1957), ''Analyticity of the Schr\"{o}dinger Scattering Amplitude and Nonrelativistic Dispersion Relations'',
\it Physical Review, \rm 107 (4): 1148-1156.\item Kontolatou, A. (1993), ''The quasi-real extension of the real
numbers'', \it Acta Mathematica et Informatica Universitatis Ostraviensis, \rm 1 (1): 27-36. \item Landau, L.D.
and Lif\v{s}its, E.M. (1982), \it Meccanica quantistica (Teoria non relativistica), \rm Roma-Mosca, IT-RUS:
Editori Riuniti-Edizioni Mir.\item Lax, M. (1950a), ''A variational method for non-conservative collisions'',
\it Physical Review, \rm 78 (3): 306-307.\item Lax, M. (1950b), ''On a well-known cross-section theorem'', \it
Physical Review, \rm 78 (3): 306-307.\item Lehmann, H. (1958), ''Analytic Properties of Scattering Amplitudes as
Functions of Momentum Transfer'', \it Il Nuovo Cimento, \rm 10 (4): 579-589.\item Love, A.E.H. (1915), ''The
Transmission of Electric Waves over the Surface of the Earth'', \it Philosophical Transactions of the Royal
Society of London, Series A, Containing Papers of a Mathematical or Physical Character, \rm 215: 105-131.\item
Ma, S.T. (1946), ''Redundant Zeros in the Discrete Energy Spectra in Heisenberg's Theory of Characteristic
Matrix'', \it Physical Review, \rm 69 (11-12) 668. \item Ma, S.T. (1947a), ''On a General Condition of
Heisenberg for the $S$ Matrix'', \it Physical Review, \rm 71 (3): 159. \item Ma, S.T. (1947b), ''Further Remarks
on the Redundant Zeros in Heisenberg's Theory of Characteristic Matrix'', \it Physical Review, \rm 71 (3):
210.\item Macdonald, H.M. (1911), ''The Diffraction of Electric Waves Round a Perfectly Reflecting Obstacle'',
\it Philosophical Transactions of the Royal Society, Series A: Mathematical, Physical and Engineering Sciences,
\rm 210: 113-144.\item Macdonald, H.M. (1914), ''The Transmission of Electric Waves around the Earth's
Surface'', \it Proceedings of the Royal Society of London, Series A: Mathematical, Physical and Engineering
Sciences, \rm 90 (615): 50-61.\item Mackey, G.W. (1978), \it Unitary Group Representations in Physics,
Probability and Number Theory, \rm San Francisco, CA: The Benjamin/Cummings Publishing Company.\item Mainardi,
F. and Pagnini, G. (2003), ''Salvatore Pincherle: the pioneer of the Mellin-Barnes integrals'', \it Journal of
Computational and Applied Mathematics, \rm 153: 331-342.\item Mandelstam, S. (1974), ''Dual-Resonance Models'',
\it Physics Reports, \rm 13 (6): 259-253.\item Marchesini, G., Ratti, S. and Scannicchio, D. (1976),
''Particelle elementari'', in: Fieschi, R. (a cura di) (1976), \it Enciclopedia della Fisica, \rm 2 Voll.,
Milano, IT: ISEDI - Istituto Editoriale Internazionale, Capitolo 12, pp. 1-92.\item Mehler, F.G. (1881),
''\"{U}eber eine mit den Kugel- und Cylinderfunctionen verwandte Funktion und ihre Anwendung in der Theorie der
Electricitätsvertheilung'', \it Mathematische Annalen, \rm 18: 161-194.\item Mellin, H. (1910), ''Abriss einer
einheitlichen Theorie der Gamma und der Hypergeometrischen Funktionen'', \it Mathematische Annalen, \rm 68:
305-337.\item Mussardo, G. (2007), \it Il modello di Ising. Introduzione alla teoria dei campi e delle
transizioni di fase, \rm Torino, IT: Bollati Boringhieri editore.\item Newton, R.G. (1964), \it The Complex
j-Plane. Complex Angular Momentum in Nonrelativistic Quantum Scattering Theory, \rm New York, NY: W.A. Benjamin,
Inc.\item Nicholson, J.W. (1907), ''A Type of Asymptotic Summation'', \it The Messenger of Mathematics, \rm 37:
84-90.\item Nicholson, J.W. (1910a), ''On the Bending of Electric Waves round the Earth'', \it The London,
Edinburgh and Dublin Philosophical Magazine and Journal of Science, \rm 19 (110): 276-278.\item Nicholson, J.W.
(1910b), ''On the Bending of Electric Waves round a Large Sphere. II'', \it The London, Edinburgh and Dublin
Philosophical Magazine and Journal of Science, \rm 20 (115): 157-172.\item Nicholson, J.W. (1911), ''On the
Bending of Electric Waves round a Large Sphere. III'', \it The London, Edinburgh and Dublin Philosophical
Magazine and Journal of Science, \rm 21 (121): 62-68.\item Omnès, R. and Froissart, M. (1963), \it Mandelstam
Theory and Regge Poles. An Introduction to Experimentalists, \rm New York, NY: W.A. Benjamin, Inc.\item
Pincherle, S. (1888), ''Sulle funzioni ipergeometriche generalizzate'', \it Atti della Reale Accademia dei
Lincei, Rendiconti della Classe di Scienze Fisiche, Matematiche e Naturali, Serie IV, \rm 4: 694-700,
792-799.\item Pincherle, S. (1965), \it Hypergeometric Functions and Various Related Problems, \rm Translation
of ''Delle funzioni ipergeometriche e di varie questioni ad esse attinenti'', \it Giornale di Matematiche di
Battaglini, \rm Vol. 32 (1894) pp. 209-291, Washington, WA: NASA - National Aeronautics and Space
Administration.\item Poincaré, H.J. (1884a), ''Sur les groupe des équations linéaires'', \it Acta Mathematica,
\rm 4: 201-311. \item Poincaré, H.J. (1884b), ''Memoiré sur les fonctions zétafuchsienne'', \it Acta
Mathematica, \rm 5: 209-278.\item Poincaré, H.J. (1903), ''Sur la Diffraction des Ondes Electriques: A Propos
d'un Article de M. Macdonald'', \it Proceedings of the Royal Society of London, \rm 72: 42-52.\item Poincaré,
H.J. (1910), ''Sur la diffraction des ondes hertziennes'', \it Rendiconti del Circolo Matematico di Palermo, \rm
XXIX: 169-259.\item Regge, T. (1959), ''Introduction to Complex Orbital Momenta'', \it Il Nuovo Cimento, \rm 14
(5): 951-976.\item Rickles, D. (2014), \it A Brief History of String Theory. From Dual Models to $M$-Theory, \rm
Berlin and Heidelberg, DE: Springer-Verlag.\item Rossetti, C. (1985), \it Elementi di teoria dell'urto, \rm
Torino, IT: Libreria Editrice Universitaria Levrotto \& Bella.\item Schwarz, J.H. (1973), ''Dual Resonance
Theory'', \it Physics Reports, \rm 8 (4): 269-335.\item Schwarz, J.H. (1975), ''Modelli a risonanza duale delle
particelle elementari'', \it Le Scienze, \rm Giugno 1975, Vol. XIV, No. 82, pp. 28-34.\item Sommerfeld, A.
(1949), \it Partial Differential Equations, \rm English Translation of the first 1947 German Edition \it
Partielle Differentialgleichungen der Physik - Vorlesungen \"{u}ber theoretische Physik, Band 6, \rm New York,
NY: Academic Press Inc., Publishers.\item Umezawa, H. and Vitiello, G. (1985), \it Quantum Mechanics, \rm
Napoli, IT: Bibliopolis, edizioni di filosofia e scienze, spa.\item Veneziano, G. (1968), ''Construction of a
Crossing-Simmetric, Regge-Behaved Amplitude for Linearly Rising Trajectories'', \it Il Nuovo Cimento, Serie A,
\rm 57 (1): 190-197.\item Watson, G.N. (1918), ''The Diffraction of Electric Waves by the Earth'', \it
Proceedings of the Royal Society of London, Series A: Mathematical, Physical and Engineering Sciences, \rm 95
(666): 83-99.\item Watson, G.N. (1944), \it A Treatise on the Theory of Bessel Functions, \rm 2nd Edition,
Cambridge, UK: Cambridge University Press.\item Whittaker, E.T. and Watson, G.N. (1927), \it A Course in Modern
Analysis. An Introduction to the General Theory of Infinite Processes and of Analytic Functions; with an Account
of the Principal Transcendental Functions, \rm Fourth Edition, Cambridge, UK: Cambridge University Press.
\end{description}

\end{document}